\documentclass[twocolumn,twocolappendix,tighten,times,astrosymb]{aastex631}

\usepackage{mathtools}
\renewcommand{\vec}[1]{\mathbf{#1}}

\shorttitle{Pair-loaded Relativistic Shocks}
\shortauthors{Gro\v selj, Sironi, \& Beloborodov}

\graphicspath{{./}{figures/}}

\begin{document}

\title{\large Microphysics of Relativistic Collisionless Electron-ion-positron Shocks}

\author[0000-0002-5408-3046]{Daniel Gro\v selj}
\affiliation{Department of Astronomy and Columbia Astrophysics Laboratory, 
Columbia University, New York, NY 10027, USA}
\correspondingauthor{Daniel Gro\v selj}
\email{daniel.groselj@columbia.edu}

\author[0000-0002-1227-2754]{Lorenzo Sironi}
\affiliation{Department of Astronomy and Columbia Astrophysics 
Laboratory, Columbia University, New York, NY 10027, USA}

\author[0000-0001-5660-3175]{Andrei M.~Beloborodov}
\affiliation{Physics Department and Columbia Astrophysics Laboratory, Columbia University, New York, NY 10027,USA}
\affiliation{Max Planck Institute for Astrophysics, D-85741 Garching, Germany}

\begin{abstract}
We perform particle-in-cell simulations to elucidate the microphysics of relativistic weakly magnetized shocks loaded with electron-positron pairs. Various external magnetizations $\sigma\lesssim 10^{-4}$ and pair-loading factors $Z_\pm \lesssim 10$ are studied, where $Z_\pm$ is the number  of loaded electrons and positrons per ion. We find the following.  (1) The shock becomes mediated by the ion Larmor gyration in the mean field when 
$\sigma$ exceeds a critical value $\sigma_{\rm L}$ that decreases with $Z_\pm$. At $\sigma\lesssim\sigma_{\rm L}$ the shock is mediated by particle scattering in the self-generated microturbulent fields, the  strength and scale of which decrease with $Z_\pm$, leading to lower $\sigma_{\rm L}$. (2) The energy fraction carried by the post-shock pairs is robustly in the range between 20\% and 50\% of the upstream ion energy. The mean energy per post-shock electron  scales as $\overline{E}_{\rm e}\propto (Z_\pm+1)^{-1}$.  (3) Pair loading suppresses nonthermal ion acceleration at magnetizations as low as $\sigma\approx 5\times 10^{-6}$. The ions then become essentially thermal
with mean energy $\overline{E}_{\rm i}$, while electrons form a nonthermal tail, extending from $E\sim (Z_\pm + 1)^{-1}\overline{E}_{\rm i}$ to
$\overline{E}_{\rm i}$. When $\sigma = 0$, particle acceleration is enhanced by the formation of intense magnetic cavities that populate the precursor during the late stages of shock evolution. Here, the maximum energy of the nonthermal ions and electrons keeps growing over the duration of the simulation. Alongside the simulations, we develop theoretical estimates consistent with the numerical results. Our findings have important implications for models of early gamma-ray burst afterglows.
\end{abstract}


\section{Introduction}

Relativistic collisionless shocks play a key role in gamma-ray bursts (GRBs), the most powerful explosions in the universe. 
The prompt GRB spectrum peaks around 1~MeV and is followed by softer afterglow emitted by the blast wave from the explosion, 
as it expands into the external medium. This external shock is weakly magnetized and
ultrarelativistic, with a Lorentz factor exceeding one hundred, and gradually decelerates with time. Its key feature is 
the ability to heat the medium to a relativistic temperature and accelerate nonthermal particles to high energies, 
which results in a broadband nonthermal afterglow radiation.

\subsection{Pair loading in external GRB shocks}

Over the past couple of decades,
relativistic collisionless shocks have been studied in detail 
using first-principles kinetic simulations. This includes, in particular,
simulations of relativistic shocks propagating in a weakly magnetized electron-ion medium (see Sec.~\ref{sec:previous}), which is expected around GRBs.
However, existing simulations do not apply to the earliest and brightest phase of the GRB afterglow, 
emitted at radii $R\lesssim 10^{17}\,$cm. At these radii, the prompt gamma-rays streaming ahead 
of the blast wave load the external medium with copious electron-positron pairs \citep{Thompson2000,Meszaros2001,Beloborodov2002}. 
The number of loaded electrons and positrons per ion, $Z_\pm$, is independent 
of the original plasma density and can be accurately calculated for any GRB 
with a known (observed) prompt gamma-ray spectrum \citep{Beloborodov2002,Beloborodov2014}. This 
calculation gives $Z_\pm> 1$ at radii $R \lesssim R_\pm\approx 10^{17}({\cal E}/10^{54}\,{\rm erg})^{1/2}$\,cm, 
where ${\cal E}$ is the isotropic equivalent of the GRB energy. At radii $R\ll R_\pm$, the pair-loading 
factor $Z_\pm$ reaches extremely high values, exceeding $10^4$, 
and  drops to $Z_\pm<1$ when the blast wave expands to $R \gtrsim R_\pm$.
In addition to the pair loading by the prompt MeV radiation, pairs 
can be created by gamma-rays emitted by the shock itself \citep{Derishev2016}.

When $Z_\pm\ll m_{\rm i}/m_{\rm e}\approx 1836$ (the proton-electron mass ratio), 
the plasma rest mass is dominated by the ions rather than pairs. On the other hand, even a modest 
$Z_\pm$ of a few can qualitatively change the shock physics, because it introduces light 
charges of both signs. This can affect the strength of magnetic fields generated in the shock and 
the mechanism of particle energization. Moreover, when the shock energy 
budget is dominated by the ions, it is important to know what fraction $\epsilon_{\rm e}$ of 
the initial ion energy will be given to the post-shock electrons 
and positrons (which can efficiently radiate) and what nonthermal tail should 
be expected in the downstream particle distribution.

Answers to these questions have strong implications for the expected early afterglow of GRBs. In particular, \citet{Beloborodov2014} proposed 
that the pair-loading factor $Z_\pm(R)$ shapes the evolution 
of early GeV emission detected in GRBs \citep{Ackermann2013}. 
Their calculations assumed that the emission is dominated by hot 
pairs with $\epsilon_{\rm e}\approx 0.3$ when $1\lesssim Z_\pm\ll m_{\rm i}/m_{\rm e}$, neglecting any nonthermal tails. 
This simple model was found consistent with observations of seven GRBs with good early GeV data, and was further 
confirmed by the optical data available for two bursts \citep{Hascoet2015}.

In order to improve confidence in models of the early GRB afterglow emission, it is 
important to constrain from first principles the relevant shock microphysics, such as the downstream pair energy fraction. To this end, we perform a set of 
particle-in-cell (PIC) simulations of collisionless weakly magnetized relativistic shocks, 
treating $Z_\pm$ as a fixed parameter for the upstream plasma. Our simulations 
are local in the sense that the scales involved are much smaller than the shock radius $R$. 
We will show below that even moderate pair-loading factors of order unity significantly affect
the structure of the shock on kinetic scales and the resulting particle acceleration, 
whereas the pair energy fraction $\epsilon_{\rm e}$ is rather insensitive to the changes in particle composition.

\subsection{Previous simulations}
\label{sec:previous}

Kinetic simulations of relativistic electron-ion and 
electron-positron 
weakly magnetized 
shocks \citep[e.g.,][]{Sironi2013, Plotnikov2018} 
demonstrate the key role of
the magnetization $\sigma$, the ratio of the upstream Poynting to kinetic energy flux.
Above a critical magnetization $\sigma_{\rm L}$ ($\sim 10^{-4}$ for electron-ion 
and $\sim 10^{-3}$ for electron-positron shocks)
the incoming flow is stopped by the Larmor gyration of the particles in the downstream compressed mean magnetic field, 
which mediates the shock transition. The 
downstream particle energy distributions are essentially thermal. 

For magnetizations $\sigma \lesssim \sigma_{\rm L}$
the shock is mediated by plasma microinstabilities, 
most notably by the Weibel (filamentation) 
instability \citep{Weibel1959, Fried1959,Medvedev1999, Silva2003, Achterberg2007a,Bret2014,Takamoto2018, Lemoine2019}. 
The Weibel instability is fueled by the anisotropy of the upstream particle momentum distribution, 
composed of the incoming background plasma and the counterstreaming beam 
of particles returning from the shock.
It acts to exponentially amplify seed magnetic fields by channeling particles into 
elongated current filaments of alternating polarity, 
which provides a positive feedback on the field perturbation. 
The filaments are elongated along the streaming direction and 
their typical thickness is comparable to the plasma skin depth.

During the nonlinear stage of the Weibel instability, incoming particles scatter off the 
self-generated turbulence, thereby isotropizing their momenta. This provides the 
mechanism that mediates the shock.
A fraction of particles is reflected back and forth across the shock 
front, gaining 
energy upon each reflection in a first-order Fermi process \citep{Blandford1987,Achterberg2001}.
Thus, Weibel mediated relativistic shocks are efficient 
particle accelerators \citep{Spitkovsky2008b, Spitkovsky2008, Martins2009, Nishikawa2009,Haugbolle2011},
In the electron-ion case, it has been also demonstrated that the incoming electrons 
are preheated to nearly 40\% of the initial ion energy before entering the downstream \citep{Spitkovsky2008,Sironi2013}. 
Essentially, the preheating eliminates the disparity between the
electron and ion plasma microscales, so that an electron-ion Weibel mediated 
shock behaves qualitatively almost as if it were composed of electrons and positrons.

\subsection{Scope of the present paper}

While shocks in electron-positron and electron-ion plasmas have been studied in detail,
only a limited number of simulations of electron-ion-positron shocks 
have been performed \citep{Hoshino1991,Hoshino1992,Amato2006,Stockem2012}, 
and all of them focused on moderate to high magnetizations, with applications 
to the termination shock of pulsar winds. 
This case is qualitatively different from the GRB blast waves that 
propagate in a very low-$\sigma$ external medium.

Here, we perform kinetic simulations of \emph{weakly magnetized}, 
relativistic pair-loaded shocks with the goal of understanding how the shock microphysics 
depends on the plasma composition. The simulations provide a fairly comprehensive view of the 
relevant parameter space, with magnetizations in the range $0\leq \sigma\leq 10^{-4}$ 
and pair-loading factors $0\leq Z_\pm\leq 12$.
Our numerical effort is complemented by
analytical estimates which help interpret the results.

The paper is organized as follows. In Sec.~\ref{sec:setup} we provide the numerical details of our shock
simulations. We first demonstrate the role of pair loading in an idealized Weibel 
unstable plasma in Sec.~\ref{sec:weibel}. This simplified model helps to interpret the main results of our shock simulations,
which are presented in Secs.~\ref{sec:structure} and \ref{sec:energetics}. Sec.~\ref{sec:structure}
shows how the shock structure changes with respect to the pair-loading factor. In Sec.~\ref{sec:energetics} we analyze the
collisionless partitioning of energy between ions and pairs, and characterize their downstream energy spectra.
The implications of our results for the early afterglow phase of GRBs are briefly discussed in 
Sec.~\ref{sec:astro}. We conclude the paper with a summary of our main results in Sec.~\ref{sec:conclusion}.

\section{Simulation setup}
\label{sec:setup}

We carried out a series of two-dimensional (2D) PIC simulations of relativistic electron-ion-positron 
shocks using the code \textsc{osiris 4.0} \citep{Fonseca2002,Fonseca2013}. 
The simulations are performed 
for various
pair-loading factors
\begin{align}
Z_{\pm} \equiv \frac{2n_{0\rm e^+}}{n_{\rm 0i}}
\end{align}
and magnetizations 
\begin{align}
\sigma \equiv \frac{B_0^2}{4\pi \gamma_0 n_{\rm 0i}m_{\rm i}c^2},
\end{align}
where $n_{\rm 0i}$ is the ion density, $n_{\rm 0\rm e^+}$ the positron 
density, $\gamma_0\gg 1$ is the Lorentz factor of the cold upstream flow, and $B_0$ is the mean shock-perpendicular magnetic field.
Subscript ``0'' refers to the upstream plasma far ahead
of the shock. All quantities are measured in the simulation frame,
in which the shocked downstream plasma is at rest. 
We consider magnetizations in the range $0\leq\sigma\leq 10^{-4}$ and pair-loading 
factors $0\leq Z_\pm \leq 12$.

To save computational resources we opt for a reduced ion-electron
mass ratio of $m_{\rm i}/m_{\rm e}=36$. We mostly focus on pair-loading factors of order unity, 
such that the ions dominate the upstream momentum even at the reduced value of the mass ratio.
Results from a simulation with $Z_\pm = 2$ and $m_{\rm i}/m_{\rm e}=100$ are included 
for reference in Appendix~\ref{app:mass_ratio}, showing good agreement with our
fiducial case $m_{\rm i}/m_{\rm e}=36$.
The upstream magnetic field $\vec B_0 = B_0\,\vec{\hat z}$ 
points out of the 2D simulation plane.\footnote{In the weakly magnetized 
relativistic regime, the out-of-plane field orientation is preferred over the
in-plane configuration, because it best captures the physics of particle acceleration \citep{Sironi2013}.}
We also initialize a motional electric 
field $\vec E_{0} = - \boldsymbol{\beta}_0\times\vec B_0$, where 
$\boldsymbol{\beta}_0 = - (1 - 1/\gamma_0^2)^{1/2}\,\vec{\hat x}$ is the initial three-velocity
of the upstream flow in units of $c$. 
The shock formation is triggered by the reflection of particles 
from a conducting wall located on the left side ($x=0$) of the 
computational
domain \citep[e.g., see][]{Spitkovsky2008,Martins2009,Sironi2013}.
The longitudinal size is chosen long enough to accommodate the 
propagating shock front until the end of the simulation.
Periodic boundaries are used in the transverse $y$
direction. The transverse size of the domain is $31.4\,d_{\rm i}$, where
$d_{\rm i} = c/\omega_{\rm pi} = (\gamma_0 m_{\rm i} c^2/4\pi n_{0\rm i}e^2)^{1/2}$ is the (upstream) relativistic ion skin depth.

The numerical details of our simulations are as follows.
We set the resolution to eight cells per pair plasma skin depth 
$d_{\rm e} = c/\omega_{\rm pe} = (\gamma_0 m_{\rm e} c^2/4\pi n_{\rm 0e}e^2)^{1/2}$, 
where $n_{0\rm e}=(Z_\pm\!+\!1)n_{\rm 0i}$ is the combined (upstream) density of electrons and positrons. 
Our time step is $\Delta t\omega_{\rm pe} = 1 / 16$. 
The calculations require significant resources, because the scale separation, 
$\omega_{\rm pe} / \omega_{\rm pi} = [(Z_\pm + 1)m_{\rm i}/m_{\rm e}]^{1/2}$, grows with 
the amount of pair loading. For instance, our largest simulation 
spans about 5400 $\times$ 295,000 grid cells and is evolved over 600,000 time steps.
Cubic spline macroparticle shapes and smoothing of the electric currents are used to 
reduce PIC noise and numerical heating. The electric current deposit is charge conserving.
An electromagnetic field solver introduced by \citet{Blinne2018} is used to mitigate 
the numerical Cherenkov instability \citep{Godfrey1974, Godfrey2013}. The upstream plasma ahead of the 
shock is introduced by a moving particle injector that is initially located next to the reflecting wall, 
but moves away from it at the speed of light as time progresses.
The injected particles are sampled from a distribution with bulk Lorentz factor $\gamma_0=50$ 
and with a thermal spread of $T_{\rm 0e^-} = T_{0\rm e^+} = T_{\rm 0i} = 4.8\times 10^{-5} m_{\rm i} c^2$.
The injected particle number is typically set to eight or twelve per cell per species.
Higher spatial resolutions and larger numbers of particles per cell were tested, indicating a qualitative and quantitative 
convergence of our results.

\section{Homogeneous beam-symmetric system}
\label{sec:weibel}

To understand how the pair enrichment affects the structure of a weakly magnetized shock, 
it is instructive to consider first an idealized periodic system, broadly resembling the
early stage of shock formation. We shall assume that the initial 
configuration consists of two symmetric, unmagnetized cold plasma shells streaming through each other.
Each of the two shells is charge and current neutral and moves with a bulk Lorentz factor $\gamma_0\gg 1$.
The ions have a total simulation-frame density $n_{\rm 0i}$ and the total
density of electrons and positrons is $n_{0\rm e} = (Z_\pm + 1) n_{\rm 0i}$. To focus on regimes
where ions dominate the energy budget we impose $Z_\pm \ll m_{\rm i} / m_{\rm e}$. 
 
The idealized configuration described above is prone to plasma streaming instabilities, 
the most prominent of which is in this context the Weibel (filamentation)
instability \citep{Weibel1959, Fried1959, Silva2003, Achterberg2007a,Kumar2015,Takamoto2018}. 
The available free energy can be also channeled into the oblique two-stream modes which are, 
unlike the Weibel instability, of the resonant type and predominantly electrostatic \citep{Bret2009,Lemoine2010}. 
These resonant modes are inhibited through Landau damping on the electrons as soon as the latter 
are heated to relativistic temperatures \citep{Bret2010a, Lemoine2011,Shaisultanov2012}. Thus, in accord with
previous works we expect the overall strength of the saturated fields to be mainly 
controlled by the Weibel instability; an assumption well supported by our numeric results presented below.

Let us consider the effect of the pair-loading parameter $Z_\pm$ on the generation of Weibel fields. 
The pair-driven instability will grow first and 
saturate on pair plasma scales, followed by the slower ion response. 
At this point, the electrons and positrons can be reasonably approximated as an isotropic, relativistically 
hot background, whereas the counterstreaming ion beams 
are still cold. The ion Weibel instability grows initially over 
the hot electron (and positron) background at a maximum rate that depends only on the ion properties.
For cold ion beams, the peak growth rate is $\Gamma \simeq \omega_{\rm pi}$,
where $\omega_{\rm pi}$ is the relativistic ion plasma frequency \citep[e.g., see][]{Achterberg2007b, Lemoine2011}. 
We will show below that, unlike the linear growth rate, the nonlinear saturation strength of ion Weibel fields depends 
strongly on $Z_\pm$ as a result of the screening of ion currents by the pair plasma background.

Saturation of the ion-driven instability proceeds as follows. The exponential Weibel field 
growth at a given beam-perpendicular
wavenumber $k$ stalls when the magnetic bounce frequency 
becomes comparable to the characteristic growth rate:
\begin{align}
 \left(\frac{e\, \beta_{x0}\delta B_k k}{\gamma_0 m_i}\right)^{1/2} \simeq \Gamma_k,
\label{eq:trapping}
\end{align}
where $\delta B_k$ is the magnetic fluctuation amplitude on scale $1/k$ and $\Gamma_k$ is the growth rate.
Condition \eqref{eq:trapping} is known
as the trapping criterion \citep{Davidson1972}. In principle, the field may be amplified 
further after the end of the linear, exponentially growing stage. A more generic but equivalent 
estimate of $\delta B_k$ can be obtained by assuming that the maximum field strength at scale $1/k$ 
is reached when all the available current has been used \citep{Kato2005,Gedalin2012}.
From Ampere's law it follows that
\begin{align}
\delta B_k \simeq 2\pi e n_{\rm 0i} / k.
\label{eq:ampere}
\end{align}
The latter leads to the same qualitative conclusion. Namely, 
as the field energy grows, short-wavelength modes saturate first, followed by ever increasing scales, up to the largest scale
that can sustain growth. 

The maximum scale over which the ion instability can grow at a rate close to the
maximum ($\sim\omega_{\rm pi}$), hereafter denoted with 
$\lambda \simeq 1/k_*$, is controlled by the electron (and positron) 
background \citep{Achterberg2007b,Kumar2015}. For a relativistically hot and
isotropic electron background, $k_*$ is estimated as
\begin{align} 
k_* d_{\rm i} \simeq \left(\frac{\omega_{\rm pi}^2}{\tilde\omega_{\rm pe}^2}\right)^{-1/3},
\label{eq:screen}
\end{align} 
where $\tilde\omega_{\rm pe}=(4\pi e^2 n_{0\rm e}/\overline\gamma_{\rm e} m_{\rm e})^{1/2}$ and $\overline\gamma_{\rm e}$ is the mean
electron (and positron) Lorentz factor. For a detailed derivation of expression \eqref{eq:screen} see 
\citet{Lyubarsky2006,Achterberg2007a}. We mention that the inhibition of field growth at 
wavenumbers $k \lesssim k_*$ originates from the screening of ion current filaments by the electrons (and positrons); 
a robust feature known to persist well 
beyond the linear stage of the instability \citep{Achterberg2007b, Ruyer2015}.

A rough estimate for the total magnetic energy at saturation can be obtained by noticing that the fluctuation amplitude
is proportional to the scale over which the field grows. Thus, the dominant contribution comes from the largest scale.
By evaluating \eqref{eq:ampere} at $k=k_*$ we are led to
\begin{align}
\epsilon_B \simeq \frac{1}{8} 
\left(\frac{\omega_{\rm pi}^2}{\tilde\omega_{\rm pe}^2}\right)^{2/3},
\label{eq:sat_weibel}
\end{align}
where $\epsilon_B \equiv \delta B^2 / 8\pi n_{\rm 0i} \gamma_0 m_{\rm i} c^2$ is the magnetic
energy fraction. Eq.~\eqref{eq:trapping} yields the same
estimate of $\epsilon_B$ for $k=k_*$ with $\Gamma_{k_*} = \omega_{\rm pi}/\sqrt{2}$ 
\citep{Achterberg2007b}.\footnote{Both \eqref{eq:trapping} 
and \eqref{eq:ampere} somewhat overpredict $\epsilon_B$ for the following reasons. 
Estimate \eqref{eq:ampere} is based for simplicity on the total available current, 
whereas the actual current is some fraction of the total. 
In the case of \eqref{eq:trapping}, we assume perfectly cold ion beams to 
estimate the growth rate, while in practice the growth rate may be reduced by the 
finite beam dispersion at the time when the largest scale 
of the instability is attained.
Rather than the precise value of $\epsilon_B$ our main interest here 
is its dependence on $Z_\pm$. For this 
purpose we find the estimate \eqref{eq:sat_weibel} sufficient.}

In order to obtain a more concrete prediction for $\epsilon_B$ and $\lambda$ we note that,
as the ion instability proceeds, electrons are heated beyond their 
initial energy of $\gamma_0 m_{\rm e}c^2$ by extracting a 
fraction $\epsilon_{\rm e}\sim 0.1$ 
from the ions \citep{Gedalin2012,Plotnikov2013,Kumar2015}. 
We therefore anticipate that reasonable estimates can be obtained provided that $\overline\gamma_{\rm e}$, 
which appears in the definition of $\tilde\omega_{\rm pe}$, takes into 
account the ion-to-electron energy transfer. This brings the 
ratio $\omega_{\rm pi} / \tilde\omega_{\rm pe}$ closer to
unity. By defining the pair energy fraction $\epsilon_{\rm e} \equiv n_{0\rm e}\overline\gamma_{\rm e}m_{\rm e}/n_{\rm 0i}\gamma_0m_{\rm i}$,
we express the mean electron Lorentz factor as
\begin{align}
\overline\gamma_{\rm e} \simeq (Z_\pm + 1)^{-1} \gamma_0
\epsilon_{\rm e}  m_{\rm i} / m_{\rm e}.
\label{eq:el_ene}
\end{align}
The explicit inverse dependence on $Z_\pm  +1$  reflects the fact that 
the energy drawn from the ions is distributed among a larger number of the light
charge carries with growing $Z_\pm$. Using relation \eqref{eq:el_ene} we can
express the ratio of the squared plasma frequencies as
\begin{align}
\frac{\omega_{\rm pi}^2}{\tilde\omega_{\rm pe}^2} = 
\frac{\overline\gamma_{\rm e} m_{\rm e}}{\gamma_0 m_{\rm i}} (Z_\pm + 1)^{-1} 
= \epsilon_{\rm e}(Z_\pm + 1)^{-2}.
\label{eq:freq_ratio}
\end{align}
This leads to the following estimates for the magnetic energy fraction and 
transverse coherence scale at saturation:
\begin{align}
\epsilon_B & \simeq \frac{1}{8} \epsilon_{\rm e}^{2/3} (Z_\pm + 1)^{-4/3}, & 
\lambda / d_{\rm i} & \simeq \epsilon_{\rm e}^{1/3} (Z_\pm + 1)^{-2/3}.&
\label{eq:predictions_weibel}
\end{align}
The estimates predict saturation of the ion Weibel instability at lower
field amplitudes and at smaller scales when the plasma is loaded with pairs.

\begin{figure}[htb!]
\centering
\includegraphics[width=\columnwidth]{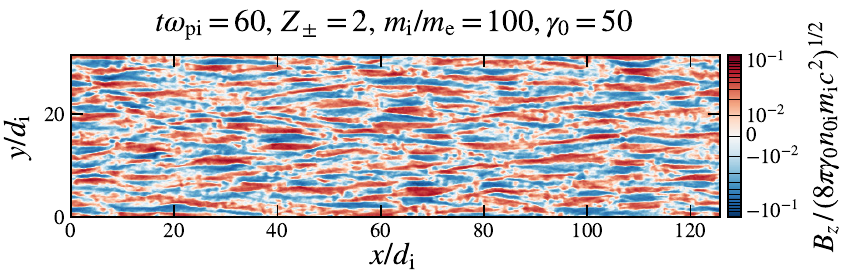}
\caption{Structure of Weibel generated magnetic fields in a moderately pair-loaded plasma.\label{fig:weibel_bz}}
\end{figure}

To test the above predictions, we perform 2D PIC simulations 
using a periodic box of size $(L_x/d_{\rm i}, L_y/d_{\rm i}) =$ (125.6, 31.4) 
and a reduced ion-electron mass ratio of 100. The initial condition consists 
of two pair-loaded cold plasma beams. The beams have opposite 
momenta and equal, spatially uniform particle densities. We evolve the system
using 32 particles per cell per species and a standard electromagnetic field solver. 
Other numerical parameters match those described in Sec.~\ref{sec:setup}
for our shock simulations.

In all runs we observe the formation of filamentary structures in the magnetic field,
as expected for the Weibel instability. An example is shown in Fig.~\ref{fig:weibel_bz}. 
A careful look at that same figure reveals also some mildly periodic
patterns along the longitudinal direction, which could be a sign of current filament 
disruption via the drift-kink instability \citep{Ruyer2018,Vanthieghem2018}.

\begin{figure}[htb!]
\centering
\includegraphics[width=\columnwidth]{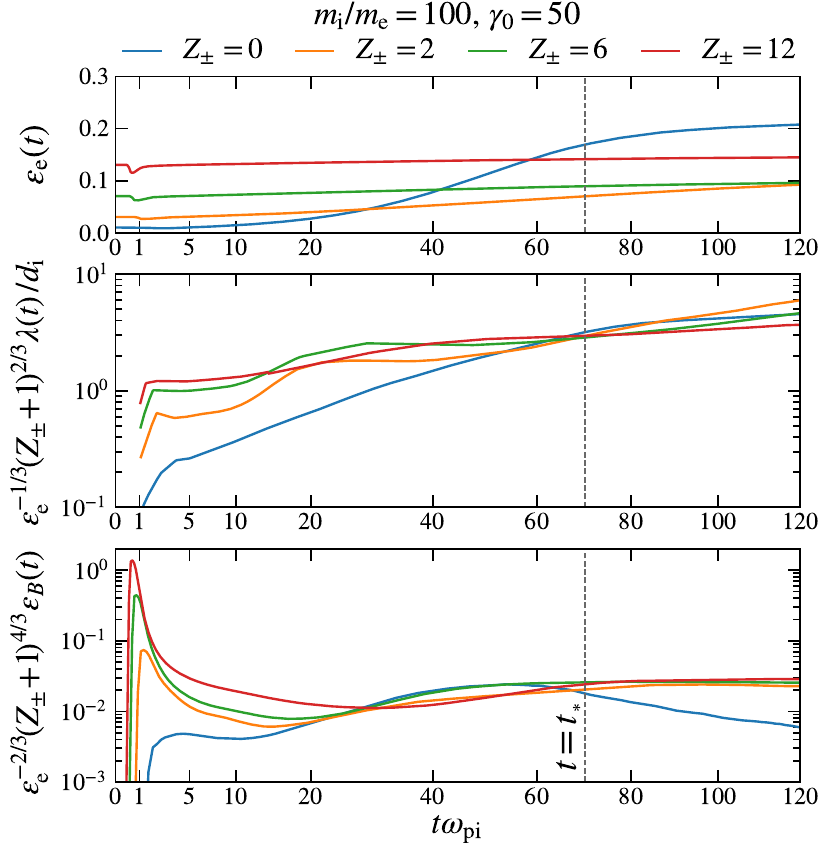}
\caption{Time evolution of the Weibel instability in a pair-loaded plasma (see main text for details).
The values of the mean magnetic energy fraction $\epsilon_B(t)$ and field coherence scale $\lambda(t)$ 
are compensated by the scaling predictions \eqref{eq:predictions_weibel}. \label{fig:weibel}}
\end{figure}

The time evolution of the system is depicted in Fig.~\ref{fig:weibel}.
The fields generated by the electron (and positron) driven instabilities saturate in a 
few tens of pair plasma times and 
decay rapidly, followed by the creation of the longer-lasting 
ion Weibel fields. We remark that the linear stage of the ion Weibel instability is not seen clearly in our setup because it 
is preceded by the faster-growing pair driven instabilities.\footnote{To see the linear evolution 
of the ion instability, we performed additional simulations with initially isotropic 
and relativistically hot pairs
and found growth rates $\Gamma\simeq\omega_{\rm pi}$, as expected for the ion Weibel instability. 
Using this same setup, we also find that the 
linear stage of the ion instability ends at around ten ion plasma times (cf.~Fig.~\ref{fig:weibel}).}

By the time $t_*\omega_{\rm pi}\approx$ 70 all simulations reach 
near maximum field strength of the ion Weibel instability (vertical dashed lines in Fig.~\ref{fig:weibel}). 
We take this as the approximate time of saturation, at which we determine the pair energy fraction $\epsilon_{\rm e}$ 
that is used to compensate the curves in the bottom two panels of Fig.~\ref{fig:weibel}.
The values of $\epsilon_{\rm e}$ are 
at near maximum around $t \approx t_*$ and increase only slightly beyond this time. 
For $Z_\pm\geq 6$, the pair energy at saturation 
hardly exceeds the initial amount at the start of the simulation, 
given by $\epsilon_{\rm e0} = (Z_\pm + 1)m_{\rm e}/m_{\rm i}$.
More specifically,  we find that the energy taken away from the ions and transferred to pairs, 
$\Delta{\epsilon}_{\rm e} \approx \epsilon_{\rm e} - \epsilon_{\rm e0}$, 
is roughly inversely proportional to $Z_\pm + 1$ (not shown).
As demonstrated below in Sec.~\ref{sec:heating}, the heating of electrons and positrons 
in weakly magnetized pair-loaded shocks turns out to 
be more efficient than what is found in this idealized setup. When 
$\epsilon_{\rm e}$ stops evolving, the screening wavenumber $k_*$ (Eq.~\eqref{eq:screen}) becomes
a constant. This sets the transverse magnetic field coherence scale $\lambda \simeq 1/k_*$
at the time of saturation of the ion instability.\footnote{Here and in 
the rest of the paper, we employ a common definition of the coherence scale \citep[e.g.,][]{Plotnikov2011} and calculate 
$\lambda$ as the power-spectrum-weighted mean of $1/k_y$, where $k_y$ is the transverse wavenumber.}

The main result of Fig.~\ref{fig:weibel} is that $\epsilon_B(t)$ and $\lambda(t)$ around the time $t\approx t_*$, 
when the ion Weibel fields reach maximum strength, 
are both nearly independent of $Z_\pm$ when compensated by the scaling predictions \eqref{eq:predictions_weibel}.
Therefore, the PIC simulations confirm that the ion Weibel fields saturate at lower amplitudes 
and at smaller scales when the plasma is enriched with pairs. 
This result has important implications for the structure of weakly magnetized relativistic shocks
with electron-ion-positron compositions, as shown in the following.

\section{Shock structure}
\label{sec:structure}

\begin{figure*}[htb!]
\centering
\includegraphics[width=\textwidth]{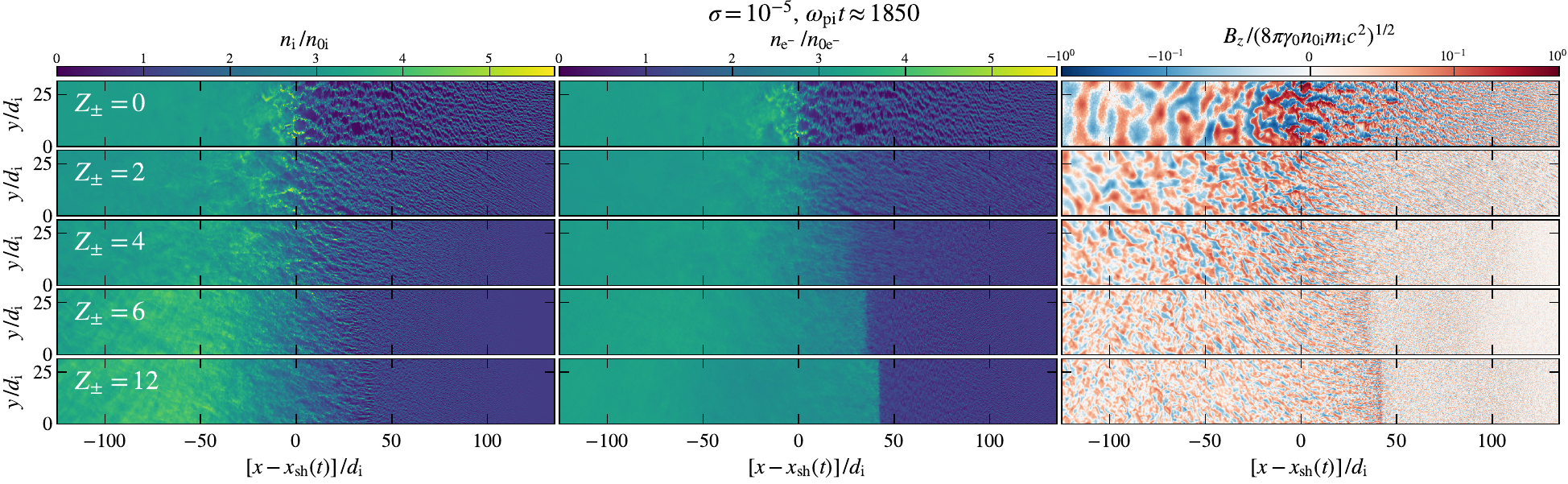}
\caption{\label{fig:transition} Shock structure 
as a function of pair-loading factor $Z_\pm$ at $\sigma=10^{-5}$. Shown 
from left to right are the ion density, 
the electron density, and the out-of-plane magnetic field. We determine
the shock position $x_{\rm sh}(t)$ as the point where the $y$-averaged 
ion density exceeds the upstream value by a factor of 2.3.}
\end{figure*}

We now turn to the shock structure as a function of the pair-loading factor. The key features
are summarized in Fig.~\ref{fig:transition}, which shows a series of simulations at 
fixed magnetization $\sigma = 10^{-5}$ and for various pair-loading factors $Z_\pm$.
The fields are visualized around $t\omega_{\rm pi} \approx$ 1850. It is evident that even 
moderate changes in the plasma 
composition significantly affect the shock structure. 
In qualitative agreement with the results of 
Sec.~\ref{sec:weibel}, the strength and scale of the self-generated magnetic turbulence drops with $Z_\pm$.
Moreover, the filamentary structure of the precursor that is characteristic of a Weibel
mediated shock fades away as pairs are introduced into the upstream plasma.

\subsection{Shock width}
\label{sec:width}

\begin{figure}[htb!]
\centering
\includegraphics[width=\columnwidth]{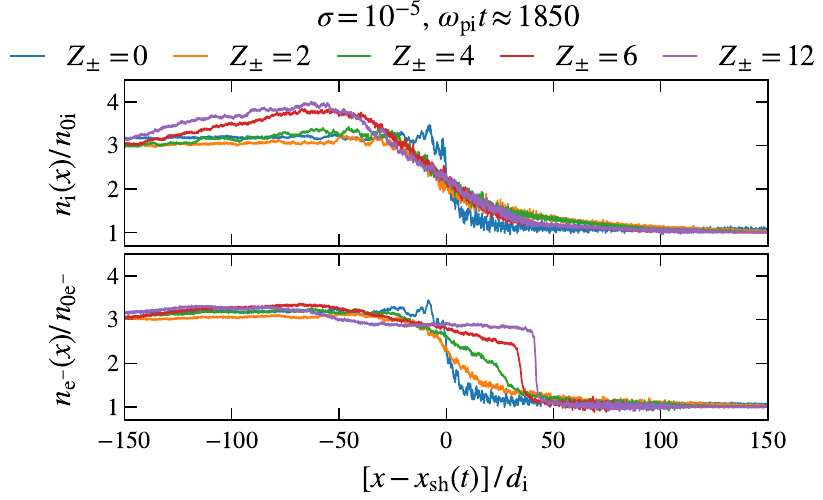}
\caption{\label{fig:den_prof} Shock density profiles for different pair-loading 
factors $Z_\pm$ at a fixed value of the magnetization $\sigma = 10^{-5}$.} 
\end{figure}

The width of the ion shock, based on the $y$-averaged ion density profile, is seen 
to broaden from about $10\, d_{\rm i}$ for $Z_\pm = 0$ 
to roughly $100\, d_{\rm i}$ for $Z_\pm \geq 6$ (Fig.~\ref{fig:den_prof}, top panel). The 
reason for the broadening is that the microturbulence becomes inefficient 
in stopping the ion flow via particle scattering (see also Sec.~\ref{sec:tracks}). 
At sufficiently large pair-loading factors ($Z_\pm \geq 6$), the width of the ion shock 
approaches the ion Larmor radius in the downstream compressed mean magnetic field, 
$R_{\rm L0}/ d_{\rm i} \simeq 1/3\sigma^{1/2} \approx 100$. Together with 
the shock structure shown in Fig.~\ref{fig:transition}, this suggests that the change 
in the plasma composition gives rise to a transition from a Weibel to a Larmor mediated 
shock at a fixed strength of the external magnetization.

The electrons and positrons (not shown) are seen to decouple 
from the ions with growing $Z_\pm$ 
and form a narrower subshock, as thin as a few $d_{\rm i}$ in width (at $Z_\pm = 12$), 
ahead of the broad ion density ramp (Fig.~\ref{fig:den_prof}, bottom panel). 
This is made possible by the fact that pairs carry lower inertia 
than the ions and are thus able to isotropize more rapidly 
than the ions when crossing the shock. In electron-ion shocks, electrostatic coupling 
prevents the formation of a narrower electron shock, even if the two species 
carry different relativistic inertia. In contrast, when the plasma is enriched with pairs, a fraction of the 
total electron charge is readily compensated by the positrons. This enables the light 
particles to decouple from the ions.

\subsection{Particle motion across the shock}
\label{sec:tracks}

\begin{figure}[htb!]
\centering
\includegraphics[width=\columnwidth]{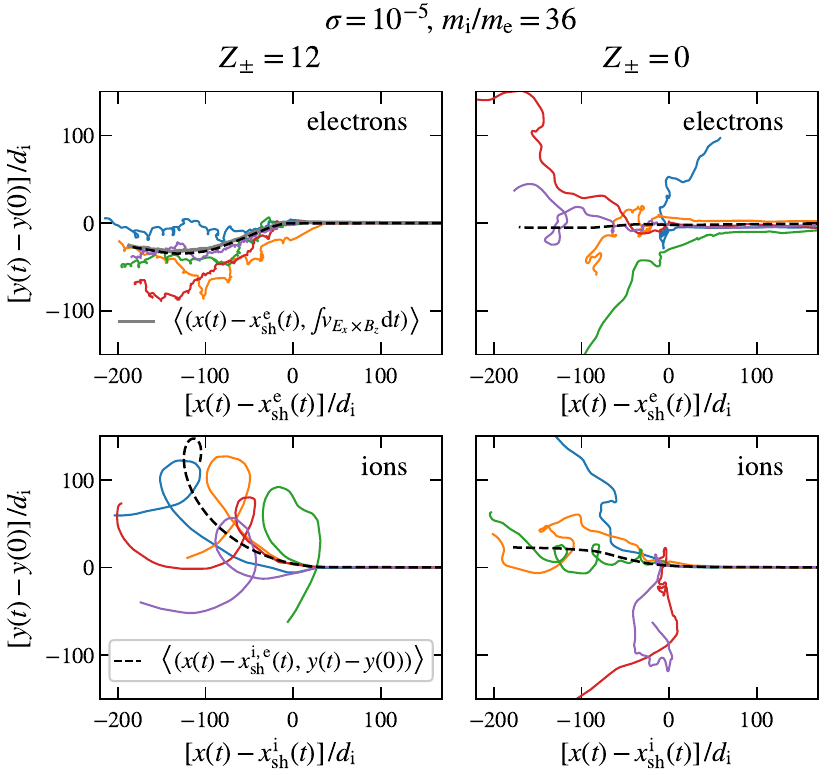}
\caption{Sample trajectories of particles crossing a $\sigma=10^{-5}$ shock for 
$Z_\pm = 12$ (left) versus $Z_\pm = 0$ (right). Dashed curves in all panels show the ensemble-averaged trajectories. 
The position $x_{\rm sh}^{\rm i,e}(t)$ denotes the point where the 
$y$-averaged density of the species shown in a given panel exceeds the upstream value by 
a factor of 2.3.
Gray curve in top left panel shows the mean transverse displacement due to the $E_x\times B_z$ drift. \label{fig:tracks}}
\end{figure}

To further elaborate on the mechanisms that mediate the shock transition at different values of 
the pair-loading parameter, we compare in Fig.~\ref{fig:tracks} the trajectories 
of particles crossing the shock for $Z_\pm = 0$ and $Z_\pm = 12$. The particle trajectories of 
the electron-ion shock ($Z_\pm = 0$) are considerably more chaotic and disperse rapidly with respect to the initial
direction of motion. The electrons acquire a significant dispersion even before entering
the shock, indicating heating in the upstream Weibel turbulence \citep{Spitkovsky2008}. 
Consistent with the phenomenology of Weibel mediated shocks, a fraction of particles performs Fermi cycles, 
scattering back and forth across the shock \citep[see also][]{Spitkovsky2008b, Martins2009}.

The motion of ions and electrons across the pair-loaded shock 
is significantly more ordered. 
The ions in particular display a very clear signature of Larmor gyration
in the compressed mean field, which mediates the shock transition. The electrons show as well 
signs of gyration, but the typical 
scale of their motions is notably smaller than that of the ions, owing to the difference 
in inertia between the species. The disparity in inertia has another consequence. 
It gives rise to a 
charge separation across the ion shock transition, which generates 
a shock-parallel electric field \citep{Lemoine2011, Lemoine2019b}. This $E_x$ field is imprinted 
onto the electron trajectories shown in the top left panel of Fig.~\ref{fig:tracks} in the 
form of a transverse $E_x\times B_z$ drift (gray curve).

\subsection{Energy fraction and scale of the magnetic fluctuations}
\label{sec:theory}

It is worth asking how the results of our shock simulations can be reconciled with the
theoretical estimates from Sec.~\ref{sec:weibel}, concerning
the saturation of ion Weibel fields in a beam-symmetric system. 
In the upstream frame of the background plasma, the maximum ion Weibel instability 
growth rate is $\Gamma_{\rm u}\simeq \omega_{\rm pb}$ \citep[e.g., see][]{Lemoine2010},
where $\omega_{\rm pb}$ is the beam plasma 
frequency of the returning ions.\footnote{If the returning 
electrons reach equipartition with the ions, as is the case for an
electron-ion Weibel mediated shock \citep{Spitkovsky2008,Sironi2013}, 
then one should strictly speaking use the combined density of 
beam ions \emph{and} electrons 
to define the beam plasma frequency \citep{Lemoine2011}. In the pair-loaded case, 
beam electrons and positrons do not contribute to the upstream turbulence as much as the beam ions, 
because they carry on average lower relativistic inertia. For simplicity, we define 
here the beam plasma frequency based on ions only.}
The latter is related to the background ion plasma frequency 
through the normalized (downstream frame) kinetic pressure of the beam ions
\begin{align}
\xi_{\rm b} \equiv 
\frac{P_{\rm b}}{\gamma_0 n_{\rm 0i} m_i c^2} = \frac{\overline\gamma_{\rm b} n_{\rm b}(\Gamma_{\rm ad} - 1)}{\gamma_0 n_{\rm 0i}},
\label{eq:xi}
\end{align}
where $\overline\gamma_{\rm b}$ is the mean Lorentz factor of the beam ions, $n_{\rm b}$ their 
density, and $\Gamma_{\rm ad}$ 
is the adiabatic index.\footnote{In a 2D geometry with 
an out-of-plane mean magnetic field, the appropriate 
adiabatic index for a relativistic gas is $\Gamma_{\rm ad}=3/2$.}
Assuming  $\overline\gamma_{\rm b} \simeq \gamma_0$, the beam plasma frequency can be expressed as 
$\omega_{\rm pb} = (4\pi e^2 n_{\rm b}/\overline\gamma_{\rm b} m_{\rm i})^{1/2} \simeq \xi_{\rm b}^{1/2} \omega_{\rm pi}$ \citep{Pelletier2017}. 
In effect, $\xi_{\rm b}$ quantifies the asymmetry of the beam-plasma system that is inherent to any realistic
shock scenario.

As the incoming plasma moves toward the shock, it experiences a growing beam 
energy density and pressure, leading to a gradual slowdown 
of the background particles over the turbulent precursor.
Instead of trying to describe the evolution over the entire precursor, 
we focus here on the generation of Weibel fields in the \emph{near} upstream, 
because this is what largely controls the nature of magnetic fluctuations 
at the shock and further downstream. The region immediately ahead of the shock 
is also where the Weibel instability plays the most prominent role, given that it is 
the most robustly growing instability once the background electrons 
become hot \citep{Lemoine2011,Shaisultanov2012,Plotnikov2013}.

In analogy with expression \eqref{eq:screen}, the growth rate of the ion beam driven instability over a background with hot electrons and cold ions 
drops below the maximum for 
transverse wavenumbers
$k\lesssim k_* \simeq \omega_{\rm pb}^{1/3}\tilde\omega_{\rm pe}^{2/3}/c$ 
\citep{Lemoine2011,Shaisultanov2012}. Using the beam parameter $\xi_{\rm b}$, the latter can be
written as
\begin{align}
k_* d_{\rm i} \simeq \xi_{\rm b}^{1/6} \left(\frac{\omega_{\rm pi}^2}{\tilde\omega_{\rm pe}^2}\right)^{-1/3}.
\label{eq:screen_shock}
\end{align}
The expression applies to transverse wavenumbers and 
as such it is frame-independent. It is obtained without taking into account 
the relative drift 
between the background species, which is appropriate for $Z_\pm = 0$ since
the background electrons and ions are in this case tightly coupled.
With a growing amount of pair loading, the motion of the background pairs 
becomes progressively more
decoupled from the ions, as discussed in Secs.~\ref{sec:width} and \ref{sec:tracks}. On this note, we mention 
that if the instability were to be driven exclusively by 
the streaming between the cold background
ions (instead of beam ions) and the hot background pairs, the screening wavenumber 
would be given by \eqref{eq:screen}, which is only marginally different from \eqref{eq:screen_shock}. 
A more detailed investigation of this aspect is left for future works.
It is also worth commenting on the possibility that the field coherence scale is ultimately determined
by the rate of current filament mergers over the 
length scale of the precursor \citep{Medvedev2005,Stockem2015,Ruyer2017}, rather than by the local 
screening effect. 
In this regard, we mention that filament merger is a slow process on scales 
exceeding the screening wavelength \citep{Achterberg2007b}, 
whereas the limited precursor length in a relativistic shock 
requires a rather fast-growing mechanism. Thus, in the relativistic case it seems reasonable
to approximate the near-upstream coherence scale with $\lambda \simeq 1/k_*$ as we do below.

Besides the coherence scale, we also require an estimate for 
the maximum available current to generate the magnetic fields.
The current filaments are produced by the response of the background plasma 
to the return particle beam, which deposits a fraction of its energy into Weibel turbulence.
We therefore identify the maximum current 
with the current of the return ion beam, 
$J_{\rm b} \simeq \xi_{\rm b}e n_{\rm 0i}c$.
Using Ampere's law based on the beam current, the maximum Weibel field strength is thus estimated as
\begin{align}
\delta B \simeq 4\pi e\xi_{\rm b} n_{\rm 0i} / k_*.
\label{eq:ampere_shock}
\end{align}

Based on \eqref{eq:freq_ratio}, \eqref{eq:screen_shock}, and \eqref{eq:ampere_shock}, 
the near-upstream magnetic energy fraction and transverse coherence scale are obtained as
\begin{align}
\epsilon_{B} & \simeq \xi_{\rm b}^{5/3}\epsilon_{\rm e}^{2/3} (Z_\pm + 1)^{-4/3}, 
& \label{eq:eps_shock}  \\
\lambda / d_{\rm i} & \simeq \xi_{\rm b}^{-1/6}\epsilon_{\rm e}^{1/3} (Z_\pm + 1)^{-2/3}.&
\label{eq:lambda_shock}
\end{align}
Compared to the symmetric 
case (Eq.~\eqref{eq:predictions_weibel}) the scalings are modified
through the addition of $\xi_{\rm b}$. 
In the near precursor of an electron-ion
Weibel mediated shock, 
typically $\xi_{\rm b} \approx 0.1$, $\epsilon_{\rm e}\approx 0.3$, 
$\epsilon_{B} \approx 0.01$, and $\lambda / d_{\rm i}\approx 1$ \citep[e.g.,][]{Sironi2013}. 
For $\xi_{\rm b}= 0.1$, $\epsilon_{\rm e}= 0.3$, and $Z_\pm = 0$, \eqref{eq:eps_shock} 
and \eqref{eq:lambda_shock} give $\epsilon_{B} \approx  0.01$ and $\lambda / d_{\rm i}\approx 1$, 
consistent with previous simulations using electron-ion
plasma compositions.

\begin{figure}[htb!]
\centering
\includegraphics[width=\columnwidth]{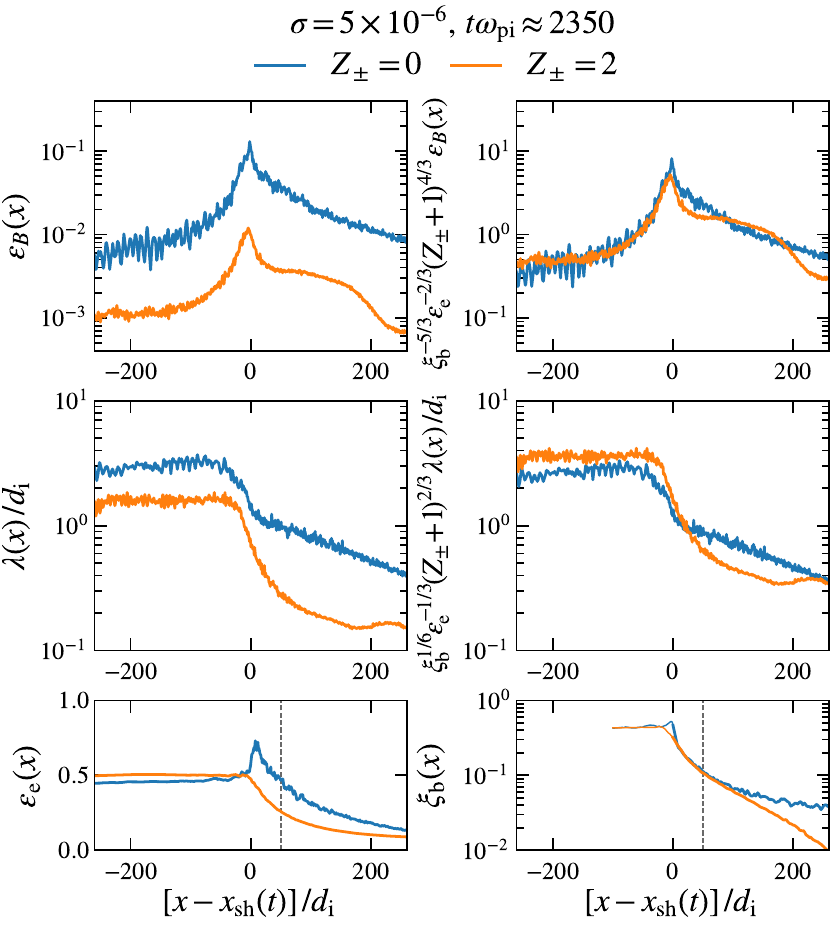}
\caption{\label{fig:profiles_compensated} Profiles of the magnetic energy fraction
$\epsilon_B(x)$ (top) and of the transverse field coherence scale $\lambda(x)$ (middle) for 
$Z_\pm = 0, 2$ at $\sigma=5\times10^{-6}$. 
The profiles are shown with no rescaling on the left. On the right, we compensate 
the curves with the scaling predictions \eqref{eq:eps_shock} and \eqref{eq:lambda_shock}. 
The bottom panels show the pair energy fraction $\epsilon_{\rm e}(x)$ (left) and the beam parameter $\xi_{\rm b}(x)$ (right).}
\end{figure}

When comparing the simulation results to the scaling 
estimates \eqref{eq:eps_shock} and \eqref{eq:lambda_shock}, one should keep in mind that the scalings are obtained 
for a steady state Weibel mediated shock with $Z_\pm \ll m_{\rm i} / m_{\rm e}$. 
In principle, the most obvious choice would be to check the predictions in the
absence of external magnetization, so that the shock is certainly Weibel mediated. 
However, as shown in Sec.~\ref{sec:unmagnetized}, the structure of a pair-loaded shock in the $\sigma=0$ limit differs 
substantially from the physics picture presented above and needs to be considered separately, owing to the creation of intense magnetic cavity structures. 
On the other hand, with increasing magnetization or pair loading the shock  moves toward the Larmor mediated regime. 
For these reasons, the predictions \eqref{eq:eps_shock} and \eqref{eq:lambda_shock} are best tested in simulations 
with a small but finite magnetization and for 
moderate pair-loading factors. This is done in Fig.~\ref{fig:profiles_compensated}, 
which shows the profiles of $\epsilon_B$ and $\lambda$ 
for $Z_\pm = 0, 2$ at $\sigma=5\times 10^{-6}$. 

The simulations shown in Fig.~\ref{fig:profiles_compensated} 
have been evolved well over 2000 $\omega_{\rm pi}^{-1}$ in order to reach a steady state. 
To make trends clearer, the curves have been shifted with respect to the (ion) shock position 
$x_{\rm sh}(t)$ and time averaged over $\approx 150\,\omega_{\rm pi}^{-1}$.
The location $x_* = 50\,d_{\rm i} + x_{\rm sh}$ 
(vertical dashed lines in bottom panels) is used as a proxy to determine the representative near-upstream values 
of $\epsilon_{\rm e}$ and $\xi_{\rm b}$ for use in 
\eqref{eq:eps_shock} and \eqref{eq:lambda_shock}.\footnote{We checked that
the predicted value of $\epsilon_B(x)$ and $\lambda(x)$ is rather insensitive to the 
precise choice of the near-upstream location where $\epsilon_{\rm e}$ and $\xi_{\rm b}$ are measured.}
We compute $\xi_{\rm b}$ by identifying ions with $\beta_x >0$ as the beam population. 
The profile of $\xi_{\rm b}$ is nearly independent of $Z_\pm$ in the
near precursor, but decays more rapidly for $Z_\pm = 2$ at larger distances, 
because the pair-loaded shock does not produce high-energy ions 
while the electron-ion shock does (see Sec.~\ref{sec:acc}). The high-energy ion beam population 
of the $Z_\pm = 0$ shock travels further upstream and seeds the microturbulence at larger distances, 
leading to a more extended region of field growth and electron preheating, as evident from the profiles of $\epsilon_{\rm e}$.

The strength of the self-generated magnetic turbulence, as quantified by $\epsilon_B(x)$ in Fig.~\ref{fig:profiles_compensated}, 
drops almost by an order in magnitude when the plasma is enriched with only a single pair per ion.
Similarly, the transverse coherence scales $\lambda(x)$ become smaller. Immediately ahead of the shock and in the downstream,
the compensation by the scaling predictions \eqref{eq:eps_shock} and \eqref{eq:lambda_shock}
nearly eliminates the difference between the results obtained 
for $Z_\pm = 0$ and $Z_\pm = 2$.
This shows that the arguments presented above offer a sensible explanation 
for why the microturbulence weakens when the upstream is loaded with electron-positron pairs. 
A central feature of the model is the screening of 
ion currents by the hot pair background, which controls the coherence scale of the
near-upstream Weibel filaments and leads to the weakening of the microturbulence with growing $Z_\pm$.

\subsection{Downstream decay of the magnetic field}
\label{sec:decay}

Up to this point, we mainly focused on the evolution of magnetic turbulence
in the near upstream of a pair-loaded shock. Downstream of the shock, the
magnetic fluctuations appear nearly static in the frame of the shocked plasma and decay via
phase mixing of the self-consistent electric currents. The results shown in 
Fig.~\ref{fig:profiles_compensated} (top panels) suggest that the magnetic 
field decay might be only moderately dependent on $Z_\pm$. Let us consider why this
might be so. 

For unmagnetized particles on scale $1/k$, the linear damping
rate of the fluctuations is estimated as
$\gamma_k \simeq (kc)^3 / \tilde\omega_{\rm pe}^2$ 
\citep[for details, see][]{Chang2008, Lemoine2015}. 
Since the typical scale of the fluctuations is comparable to $\lambda$, it is instructive to 
evaluate $\gamma_k$ for $k \sim 1/\lambda$.
Using \eqref{eq:freq_ratio} and \eqref{eq:lambda_shock}, this gives
\begin{align}
\gamma_{\!1\!/\!\lambda} \sim \omega_{\rm pi} (\omega_{\rm pi} / \tilde\omega_{\rm pe})^2 (\lambda / d_{\rm i})^{-3} 
\sim \xi_{\rm b}^{1/2}\omega_{\rm pi}.
\label{eq:damping}
\end{align}
For simplicity, we have ignored the fact that the scale $\lambda$ is estimated in the near precursor, 
whereas the damping rate concerns the downstream fluctuations.
According to this crude estimate, the damping rate on scale $\lambda$ 
depends on $Z_\pm$ only implicitly via $\xi_{\rm b}$.
Therefore, it seems possible that the overall rate of magnetic field 
decay is indeed only weakly dependent on $Z_\pm$ (as long as $\xi_{\rm b}$ does not change).
A definite answer to this question requires simulations with $Z_\pm\gg 1$ 
evolved over several thousands of ion plasma times, 
which is computationally prohibitive at present. 
From a theoretical perspective, a more complete treatment would have 
to consider the evolution of the entire magnetic field spectrum 
and possible modifications of the damping due to particle trapping 
and nonthermal features in their energy 
distribution \citep{Chang2008, Keshet2009, Lemoine2015}.
We defer a detailed investigation of these aspects to future studies.

\subsection{Critical magnetization for a Larmor mediated shock}
\label{sec:transition}

We have shown that, as the plasma is loaded with pairs, a weakly magnetized Weibel shock is 
transformed into one which is essentially Larmor mediated.
From the simulations we can infer that the critical magnetization $\sigma_{\rm L}$, 
required for the ion shock to become Larmor mediated, roughly scales as 
$\sigma_{\rm L}\propto(Z_\pm + 1)^{-1}$. 
For $Z_\pm = 2$ and $Z_\pm = 0$ we find the transition 
near $\sigma_{\rm L}\approx 3\times 10^{-5}$ and $\sigma_{\rm L}\approx 10^{-4}$ (not shown), respectively, the
latter being consistent with earlier simulations of electron-ion shocks \citep[e.g.,][]{Sironi2013}.
For $Z_\pm = 6$ we infer $\sigma_{\rm L}\approx 10^{-5}$ based on the shock structure shown
in Figs.~\ref{fig:transition} and \ref{fig:den_prof}. 

To obtain a prediction for the scaling of $\sigma_{\rm L}$ one should 
determine when the motion of the background ions across the shock becomes 
dominated by the mean field as opposed to random scattering in the fluctuating fields. 
A tentative scaling broadly consistent with our simulations can be obtained by adopting 
the scattering frequency derived by \citet{Lemoine2019b}, 
appropriate for particles that become 
trapped in the upstream Weibel filaments. The scattering frequency of 
the trapped background ions, normalized to their Larmor frequency in the mean field $\Omega_{\rm L0}$,  
is estimated as $\nu_{\rm scat} / \Omega_{\rm L0} \sim (\epsilon_B / \sigma)^{1/2} \lambda / \ell_\parallel$,
where $\ell_{\parallel}$ is a characteristic longitudinal scale of the filaments
\citep{Lemoine2019b}.
The mean field dominates the transport when $\Omega_{\rm L0} \gtrsim \nu_{\rm scat}$. Using \eqref{eq:eps_shock}, 
there follows the estimate
\begin{align}
\sigma_{\rm L} \sim 
(\lambda / \ell_\parallel)^2 \epsilon_B \sim (\lambda / \ell_\parallel)^2 
\xi_{\rm b}^{5/3}\epsilon_{\rm e}^{2/3}(Z_\pm + 1)^{-4/3}.
\end{align}
For typical values $\lambda / \ell_\parallel\approx 0.1$, $\xi_{\rm b}\approx 0.1$, $\epsilon_{\rm e}\approx 0.3$, 
this yields $\sigma_{\rm L} \sim 10^{-4}\times(Z_\pm + 1)^{-4/3}$,
which is in reasonable agreement with our simulations. For reference, the ion
Larmor radius in the rest frame of the upstream Weibel filaments 
\citep[``Weibel frame,''][]{Pelletier2019}
is  $R_{\rm L,w} / d_{\rm i} \sim \epsilon_B^{-1/2} \gamma_{\rm w}\gamma_{\rm part,w} / \gamma_0$,
where $\gamma_{\rm w}$ is the Lorentz factor of the 
Weibel frame (with respect to the downstream) 
and $\gamma_{\rm part,w}$ is a typical Lorentz factor of a particle in this frame.
From \eqref{eq:ampere_shock} and \eqref{eq:lambda_shock} it follows 
that $R_{\rm L,w}/\lambda \sim \xi_{\rm b}^{-2/3}
\epsilon_{\rm e}^{-2/3}(Z_\pm + 1)^{4/3} \gamma_{\rm w}\gamma_{\rm part,w}/\gamma_0$.
A reasonable choice of parameters ($\gamma_{\rm w}$ and $\gamma_{\rm part,w}$ both mildly relativistic, 
$\gamma_0\approx 50$, $\xi_{\rm b}\approx 0.1$, and $\epsilon_{\rm e}\approx 0.3$) gives $R_{\rm L,w}/\lambda \sim (Z_\pm + 1)^{4/3}$. 
This implies that the near-upstream background 
ions are marginally trapped ($R_{\rm L,w}\sim \lambda$) in the Weibel filaments when the composition is electron-ion 
and become progressively less magnetized with growing $Z_\pm$. In practice, the trapping regime may 
as well extend up to $Z_\pm$ of a few, given that the ions are concentrated at the shock 
in small-scale density filaments, surrounded by locally 
intense fields with amplitudes above the typical fluctuation strength (see Fig.~\ref{fig:den_filaments}).

If instead the background ions are unmagnetized, the usual estimate for the critical magnetization
gives $\sigma_{\rm L} \sim \epsilon_B^2(\lambda / d_{\rm i})^2$
\citep[e.g., see][]{Vanthieghem2020}. Together with \eqref{eq:eps_shock}
and \eqref{eq:lambda_shock}, this translates into $\sigma_{\rm L}\sim 10^{-4}\times(Z_\pm + 1)^{-4}$
for typical values of $\xi_{\rm b}$ and $\epsilon_{\rm e}$.
The latter is inconsistent with our numerical results
for pair-loading factors of order unity,
but may become relevant in high-$Z_\pm$ shocks
with realistic ion-electron mass ratios, such that $1\ll Z_\pm\ll m_{\rm i} / m_{\rm e}$; 
a regime currently inaccessible to long-duration PIC simulations.

It should be noted that the the transport of the background plasma over the
precursor of a pair-loaded shock warrants further investigation 
beyond the scope of the present work. One aspect worth mentioning concerns 
the slowdown of the background plasma under the influence of 
the perpendicular current driven instability \citep{Lemoine2014}, which is not considered
in our scaling estimates, but may play an important role in bridging the 
gap between the Weibel and Larmor mediated shock regimes.\footnote{To our knowledge, 
the perpendicular current driven instability was so far studied only in weakly magnetized pair 
plasma shocks \citep{Lemoine2014, Plotnikov2018}. Its role in shocks with different 
particle compositions is not well understood at present.}

\subsection{Unmagnetized limit}
\label{sec:unmagnetized}

\begin{figure*}[htb!]
\centering
\includegraphics[width=\textwidth]{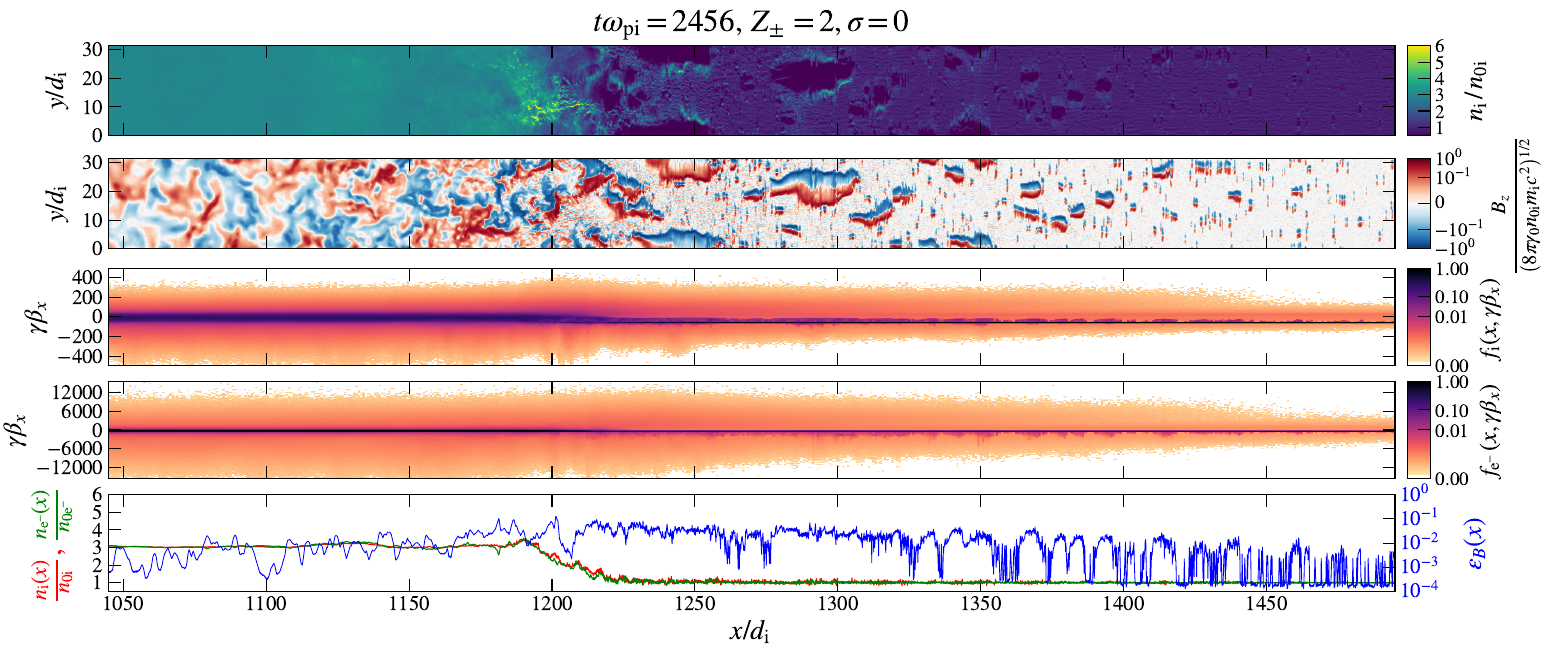}
\caption{Late-time structure of a pair-loaded shock in the absence of external magnetization ($\sigma = 0$). Shown from 
top to bottom are the ion density, the out-of-plane magnetic field, the longitudinal ion and electron
phasespace, and the $y$-averaged particle density and magnetic energy fraction.
\label{fig:unmagnetized}}
\end{figure*}

So far, we discussed the regime of small yet finite external magnetization, probing typical 
values of the order of $\sigma\sim 10^{-5}$. We showed that the self-generated microturbulence weakens as the 
plasma is loaded with pairs. As it turns out, this trend cannot be extrapolated to arbitrarily low $\sigma$, because 
the unmagnetized limit exhibits features qualitatively different from the weakly magnetized case. 

In Fig.~\ref{fig:unmagnetized} we show the late-time structure of an unmagnetized 
shock with moderate pair loading.\footnote{An animation showing the shock time 
evolution is available online at: \url{https://youtu.be/vHnX1n-s90Y}.} 
In this case, 
the precursor is filled with intense magnetic structures with near equipartition fields strengths. The 
structures are born out of Weibel filaments, forming cavities in the background plasma density. These 
cavities are filled with intense magnetic fields (locally as high as $\epsilon_B\sim 1$), the amplitude of which shows no apparent 
dependence on $Z_\pm$.\footnote{More specifically, we also performed 
simulations for $\sigma=0$ and $Z_\pm=4,\,6$ up to $t\omega_{\rm pi}\approx 1500$
and found no clear dependence of the local cavity field amplitude on $Z_\pm$.} 
As time progresses, the structures grow and merge, reaching scales up to several ion skin depths in size. 
In the long-time regime, the magnetic cavities penetrate
toward the near precursor and build up an intense magnetic barrier at the shock with 
a mean magnetic energy fraction of the order of $\langle\epsilon_B\rangle\sim 0.1$. 
As the particles scatter off the intense magnetic fields, their nonthermal acceleration becomes more 
efficient (see Sec.~\ref{sec:acc}). 
The limit of an unmagnetized pair-loaded shock is therefore different from the weakly magnetized regime, 
in a sense that a moderate pair enrichment does not lead to a reduced efficiency of particle scattering.

Similar structures have been previously observed in a variety of streaming unstable configurations, 
ranging from precursors of relativistic 
electron-ion shocks \citep{Naseri2018} to 
simulations of laser-plasma experiments \citep[e.g.,][]{Honda2000},
including laser-driven shocks \citep{Fiuza2012,Ruyer2015b}. More recently, the 
magnetic cavities were analyzed in simulations of relativistic 
beam-plasma instabilities \citep{Peterson2021, Bresci2021, Peterson2021b}. 
These authors showed that the growth of magnetic cavities is essentially driven by 
the relativistic beam electrons, streaming over an electron-ion 
or electron-ion-positron background. 
In particular, \citet{Peterson2021} interpret the growth of the magnetic cavities as a secondary 
nonlinear instability of Weibel filaments. In their model, the secondary instability saturates either 
when the beam electrons become trapped in the cavity or when the background ions are accelerated in 
the upstream rest frame to relativistic velocities, such that they neutralize the electron beam current. 
The high Lorentz factor of our simulated shock ($\gamma_0 = 50$) 
ensures that the relativistic inertia of the beam electrons 
exceeds $m_{\rm i}c$ ($= 36\, m_{\rm e}c$) in the upstream frame, 
and therefore the backround ions neutralize the current of the beam electrons before the latter become trapped. 
In our notation, the saturation strength obtained by \citet{Peterson2021} then becomes 
$\epsilon_B = \delta B^2 / 8\pi n_{\rm 0i} \gamma_0 m_{\rm i} c^2 \sim\alpha_{\rm u}$, where $\alpha_{\rm u}$ is the ratio between the beam electron and background ion density, measured in the upstream frame of the background ions. 
In the far precursor $\alpha_{\rm u}\ll 1$, but the ratio $\alpha_{\rm u}$ grows as the upstream plasma is 
advected closer to the shock and experiences a growing electron beam density. 
In the simulation shown in Fig.~\ref{fig:unmagnetized}, the fields reach 
$\epsilon_B \sim 1$ locally at the cavity, 
implying that effectively $\alpha_{\rm u}\sim 1$ where the amplitude saturates. 
In our present understanding, the 
key feature that enables the generation of equipartition field strengths is the fact that the 
background plasma is evacuated from the cavity. As a result, the screening effect that otherwise limits the field 
growth (see Secs.~\ref{sec:weibel} and \ref{sec:theory}) is inhibited, because there are hardly any background particles left to screen the current inside the cavity.

In the simulation depicted in Fig.~\ref{fig:unmagnetized}, the structure of the shock is still evolving even
at relatively late times. This naturally prompts the question about the ultimate fate of the magnetic cavities in
the long-time limit. Given that the structures always appear with the same magnetic field polarity, it is
evident that an asymmetry in the inertia of the different species is a necessary condition for the 
cavity generation \citep{Bresci2021,Peterson2021b}. Such asymmetry is naturally present in a pair-loaded 
shock, both for the incoming as well as the returning beam particles. For the latter, we remind that the pairs 
are heated below energy equipartition with the ions when $Z_\pm\gtrsim 1$ (see Sec.~\ref{sec:heating}). Whether this is in 
fact a sufficient condition for sustained cavity generation should be investigated further.

It should be mentioned that we observe the cavities also in our electron-ion shock simulations, as well as in pair-loaded shocks
with a low but finite $\sigma\lesssim 3\times 10^{-5}$. The key difference from the simulations for $\sigma=0$ is that the cavities are rather 
transient in nature. At finite magnetizations, the cavities appear at relatively early times, following the initial 
reflection of plasma from the simulation wall, and typically remain confined to within the far upstream without growing to large size. 
After this initial transient, the simulations at finite values of $\sigma$ approach a steady state, apparently free from the magnetic cavities.
In electron-ion simulations at $\sigma = 0$,  
we as well observe fewer cavities, but larger in size, as time progresses \citep[see also][]{Naseri2018}. 
This could potentially indicate that the near energy equipartition between the returning beam electrons 
and ions is limiting the cavity production. On the other hand, the evolution of the structures in our 
electron-ion $\sigma=0$ simulations might be as well affected by numerical limitations 
(e.g., the production of the cavities could be constrained by the limited width of the simulation box). 
Additional numerical experiments, beyond the scope of this work, are needed to clarify this aspect.

\subsection{Range of applicability of the unmagnetized limit}
\label{sec:unmag_disc}

We showed that the limit of an unmagnetized pair-loaded shock
differs from the regime of weak but finite external magnetization.
It is worth asking how low should $\sigma$ be for the shock to be considered unmagnetized. 
A common criterion found in literature is based on the requirement that the 
upstream residence time of the returning beam particles is controlled by scattering in the self-generated fields, 
rather than by the gyration in the mean upstream magnetic field. This amounts to 
$\sigma \lesssim \sigma_{\rm um} \sim \gamma_{\rm p}^{-2}\epsilon_B^2 (\lambda / d_{\rm i})^2$
for beam particles with typical energy $E\sim \gamma_0 m_{\rm i}c^2$, where 
$\gamma_{\rm p}$ is the Lorentz factor of the incoming 
background plasma in the downstream frame \citep[e.g., see][]{Lemoine2010, Lemoine2014}.\footnote{In contrast 
to the estimates for $\sigma_{\rm L}$ (Sec.~\ref{sec:transition}), which concern the near-upstream incoming background ions,
the expression for $\sigma_{\rm um}$ applies to the returning beam particles ahead of the shock 
\citep[this brings in the $\gamma_{\rm p}^{-2}$ factor, see][]{Lemoine2010,Lemoine2014}.}
Assuming that, regardless of $\gamma_0\gg 1$, the background plasma decelerates from the far precursor to 
a typical bulk Lorentz factor $\gamma_{\rm p}\sim 10$, and that owing to the magnetic cavity generation
we have on average $\epsilon_B\sim 0.01$ and $\lambda\sim d_{\rm i}$, we obtain $\sigma\lesssim \sigma_{\rm um} \sim 10^{-6}$. 
This upper limit is consistent with our simulations, showing explicitly that magnetizations as low as $\sigma\approx 5\times 10^{-6}$ are 
too high for the shock to be considered unmagnetized. In our present understanding, it is even 
more likely that the unmagnetized limit requires $\sigma\ll 10^{-6}$. Moreover, it is possible that $\sigma_{\rm um}$ depends on $Z_\pm$. 
Very long-duration simulations at extremely low but finite $\sigma\lesssim 10^{-6}$ are required to further constrain this critical value.

\section{Energy partitioning and particle acceleration}
\label{sec:energetics}

So far, we focused on the kinetic-scale structure of a relativistic shock enriched with electron-positron pairs. 
Now, we discuss how the shock redistributes the incoming kinetic energy 
among the ions and pairs in the post-shock plasma.

\subsection{Pair energy fraction}
\label{sec:heating}

\begin{figure}[htb!]
\centering
\includegraphics[width=\columnwidth]{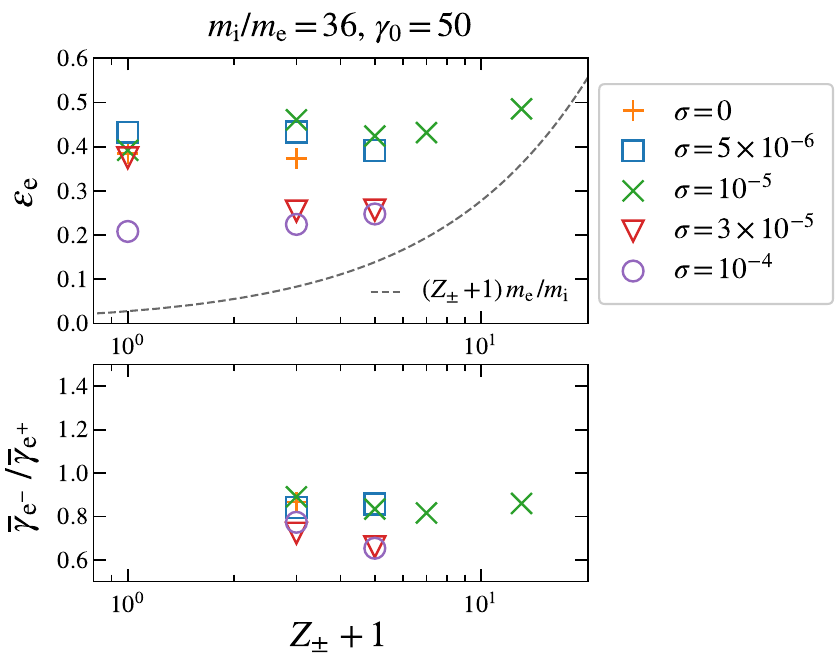}
\caption{\label{fig:pair_frac} Downstream pair energy fraction (top) 
and the electron-positron mean energy ratio (bottom). Dashed curve in 
the top panel indicates the far 
upstream energy content of the pairs.}
\end{figure}

For accurate modeling of the radiation emission, it is important to determine what fraction of energy is drawn from the ion reservoir 
and transferred to the pairs during their passage across the shock. We quantify this energy exchange with 
the pair energy fraction 
$\epsilon_{\rm e} = (Z_\pm + 1)\overline{\gamma}_{\rm e} m_{\rm e} / \gamma_0 m_{\rm i}$, 
which we measure downstream of the shock (Fig.~\ref{fig:pair_frac}). The measurements 
are obtained in a slice between $-150$ and $-100\,d_{\rm i}$ behind the ion
shock, around the time $t\omega_{\rm pi} \approx $ 1650. Quantitatively similar results are obtained at later times. 
As shown in Fig.~\ref{fig:pair_frac}, the pair energy fraction 
is robustly in the range between 20\% and 50\% over the entire range of magnetizations 
considered. Higher magnetizations (comparable to $\sigma\sim 10^{-4}$) 
yield somewhat lower pair energy fractions, around 20\%, compared to the lowest $\sigma$ range with $\sigma \lesssim 10^{-5}$, 
where the values of $\epsilon_{\rm e}$ are scattered around 40\%.

The mean energy per particle, $\overline{E}_{\rm e} = \overline{\gamma}_{\rm e}m_{\rm e}c^2$, 
is obtained directly from the definition of $\epsilon_{\rm e}$ as
\begin{align}
\overline{E}_{\rm e}= \epsilon_{\rm e}(Z_\pm + 1)^{-1}E_{0\rm i},
\end{align}
with $0.2\lesssim \epsilon_{\rm e}\lesssim 0.5$ and $E_{0\rm i} = \gamma_0 m_{\rm i}c^2$. Therefore, with 
increasing $Z_\pm$ the post-shock pairs become cooler. Their mean energy scales
approximately as $\overline{E}_{\rm e}\propto (Z\pm + 1)^{-1}$. Apart from the pair energy fraction, 
we also determine the electron-positron energy ratio (Fig.~\ref{fig:pair_frac}, bottom panel), which lies between 60\% and 90\%, 
regardless of the precise value of $Z_\pm$. 

\begin{figure}[htb!]
\centering
\includegraphics[width=\columnwidth]{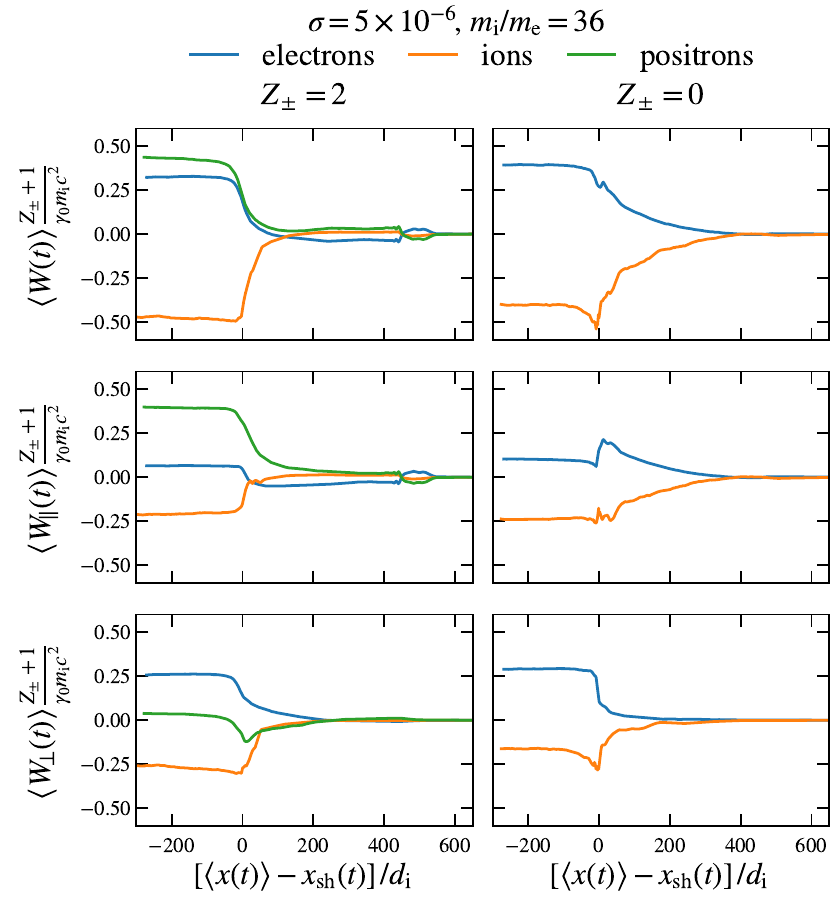}
\caption{\label{fig:work} Work by the electric field on a tracked set of background particles as they stream toward and across the shock.
Shown from top to bottom are the total work $W(t) = q\int\vec E(t)\cdot \vec v(t){\rm d}t$, the work due to the 
longitudinal $W_\parallel(t) = q\int E_x(t)v_x(t){\rm d}t$, 
and transverse field $W_\perp(t) = q\int E_y(t)v_y(t){\rm d}t$.}
\end{figure}

It is worth elaborating further on the physics of the electron and positron heating.
In Fig.~\ref{fig:work} we show the mean work done by the electric field on a set of tracked particles 
(ions, electrons, and positrons) at $\sigma=5\times10^{-6}$ for $Z_\pm = 0,\, 2$ from the far upstream, across the shock and 
into the downstream. We calculate separately the work done by the longitudinal ($E_x$) and transverse ($E_y$) 
electric field. Although the post-shock electrons are heated to nearly the same temperature as 
the positrons, we find that the positrons gain most of 
their energy by interacting with the shock-parallel $E_x$ field, 
whereas the electrons primarily receive energy from the transverse $E_y$ field. 
The ions lose energy through the interaction with both $E_x$ and $E_y$. It is also interesting that
most of the energization for $Z_\pm = 2$ occurs relatively close to the shock (within a distance of about 50 $d_{\rm i}$), 
compared to the electron-ion case. 

\begin{figure}[htb!]
\centering
\includegraphics[width=\columnwidth]{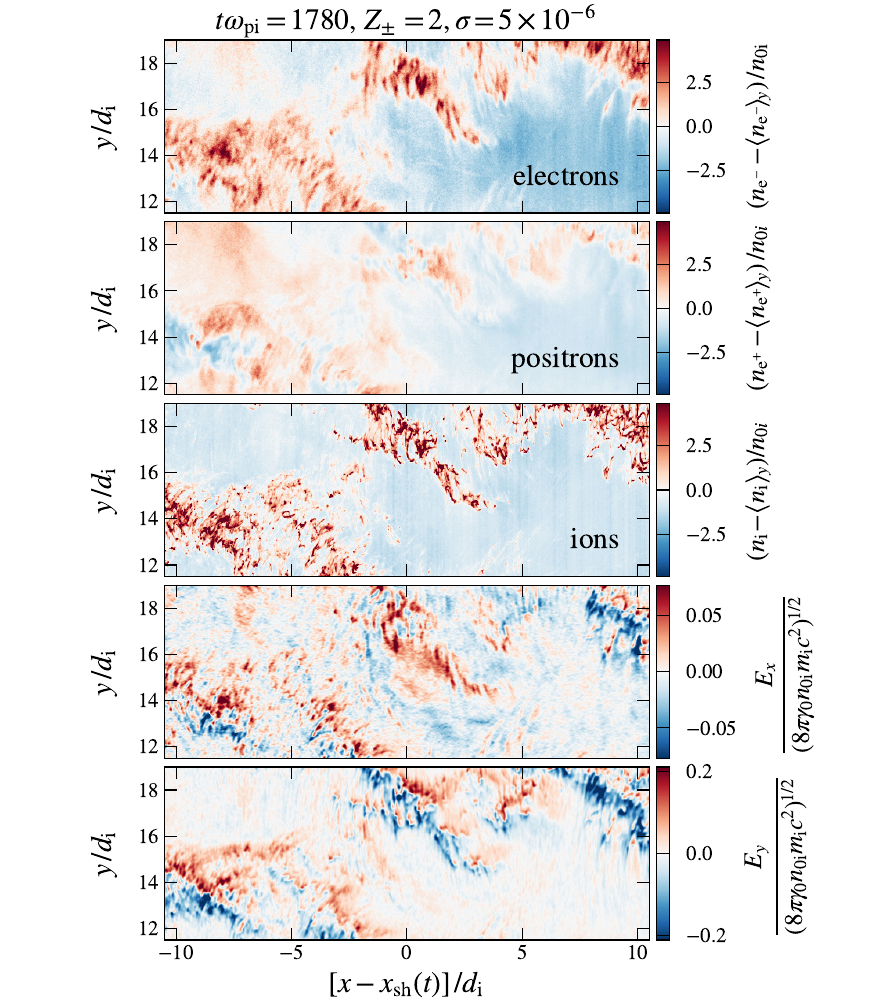}
\caption{\label{fig:den_filaments}Small-scale structure of the shock transition region in a moderately
pair-loaded plasma ($Z_\pm = 2$). Shown are the particle density fluctuations around the mean (top three panels) 
and the electric fields (bottom two panels).}
\end{figure}

The different mechanisms of electron and positron heating are related to strong ion density inhomogeneities 
near the shock transition. As shown in Fig.~\ref{fig:den_filaments}, the ions near the shock transition 
form small-scale density structures with intense fluctuations around the mean. The electrons are drawn toward these structures 
as they try to compensate the ion space charge, but their density fluctuations 
appear more diffuse due to thermal effects. 
The positrons, on the other hand, are repelled away 
from the most intense ion density fluctuations. The $E_x$ and $E_y$ electric fields (bottom two panels in Fig.~\ref{fig:den_filaments}) 
are correlated with the ion 
density inhomogeneities, and therefore the electrons receive the work by the electric field in a 
qualitatively different way than the positrons.

\subsection{Particle acceleration}
\label{sec:acc}

Previous works have shown that relativistic shocks propagating into 
an electron-ion or electron-positron medium give rise to efficient
particle acceleration via the first-order Fermi process \citep{Achterberg2001}, 
provided that the external magnetization is sufficiently weak \citep{Spitkovsky2008b,Martins2009,Sironi2013,Plotnikov2018}.
This maximum magnetization is determined by the requirement that 
the particle scattering in the microturbulence beats the motion in the compressed mean magnetic field that 
tries to advect the particles away from the shock toward the downstream \citep{Pelletier2009}.
Let us consider the implications of the argument for a pair-loaded relativistic shock.

\begin{figure*}[htb!]
\centering
\includegraphics[width=\textwidth]{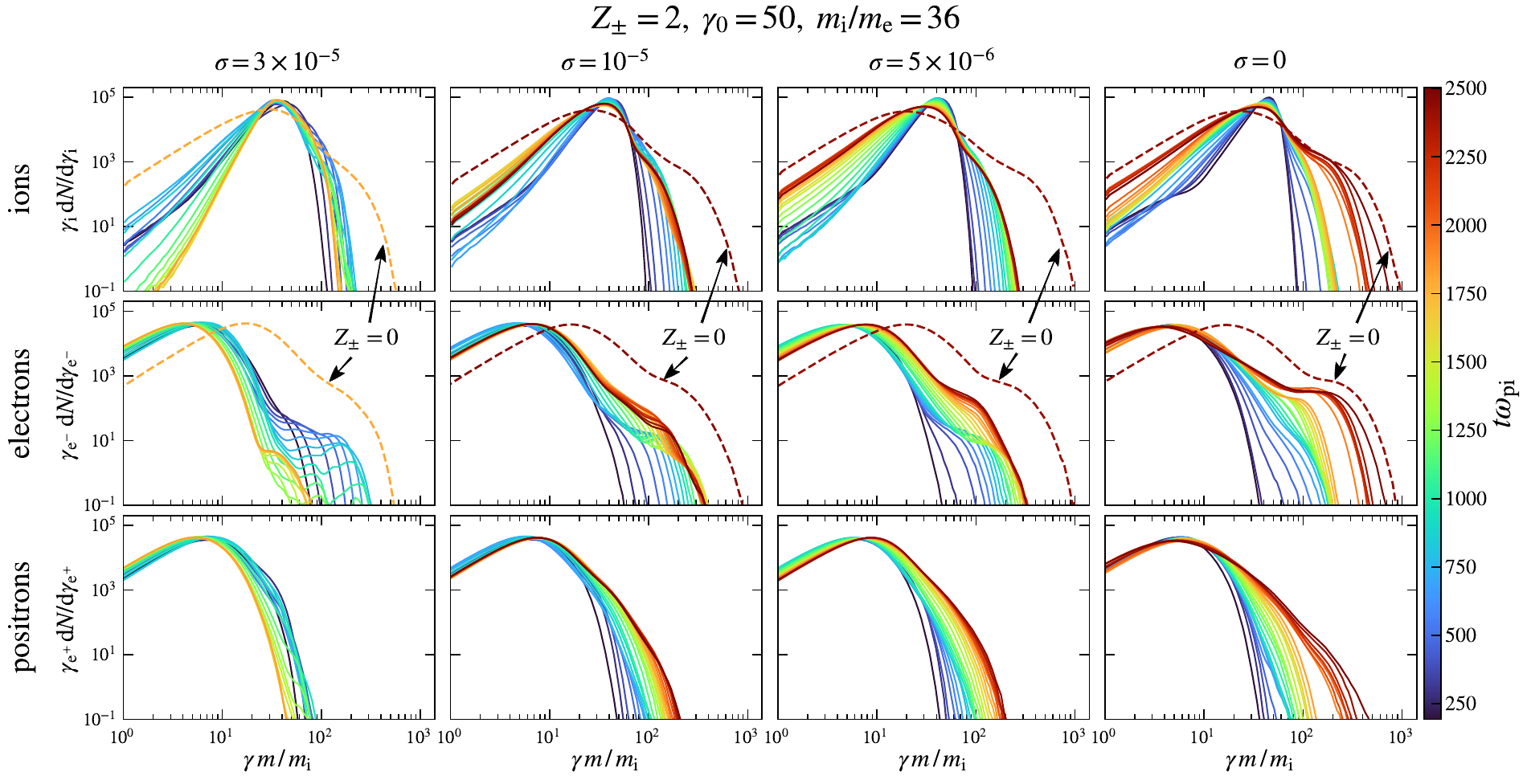}
\caption{\label{fig:spectra_Z2}Energy spectrum evolution at fixed $Z_\pm = 2$ and for various magnetizations 
$\sigma$ in a slice between $-150$ and $-100\,d_{\rm i}$ behind the ion shock. Different colors are used to represent the simulation time. Dashed curves show the 
late-time spectra from electron-ion simulations ($Z_\pm = 0$).}
\end{figure*}

As appropriate for Fermi acceleration, we shall consider small-angle random scatterings of 
untrapped particles with $R_{\rm L}\gtrsim\lambda$, where $R_{\rm L} \simeq E / e\delta B$ 
is the Larmor radius of a particle with energy $E$ in the fluctuating field $\delta B$. 
In the near downstream, a crude estimate based on \eqref{eq:ampere_shock} and \eqref{eq:lambda_shock} 
gives  $R_{\rm L} / \lambda \sim (1/3)\,\xi_{\rm b}^{-2/3}\epsilon_{\rm e}^{-2/3} (Z_\pm + 1)^{4/3} (E/\overline{E}_{\rm i})$, when
the energy $E$ is compared to the mean of the 
ion distribution $\overline{E}_{\rm i}\sim E_{\rm 0i}$, 
and $R_{\rm L} / \lambda \sim (1/3)\,\xi_{\rm b}^{-2/3}\epsilon_{\rm e}^{1/3} (Z_\pm + 1)^{1/3}(E/\overline{E}_{\rm e})$ if the mean electron energy $\overline{E}_{\rm e}$ is used instead.\footnote{The factor of $1/3$
accounts for the magnetic field compression at the 2D relativistic shock.} 
Assuming $\xi_{\rm b}\approx 0.1$ and $\epsilon_{\rm e}\approx 0.3$, we find that the suprathermal ions and electrons with energies exceeding the mean (for each species) by factors of a few
are always unmagnetized ($R_{\rm L}\gtrsim\lambda$), and even more so at higher $Z_\pm$.

For the unmagnetized suprathermal particles, the downstream scattering frequency 
is estimated as $\nu_{\rm scat} \simeq \lambda c /R_{\rm L}^2$ \citep[e.g.,][]{Plotnikov2011}.
Scattering prevails when $\nu_{\rm scat} \gtrsim \Omega_{\rm L0}\simeq eB_0c/E$,
thereby enabling the particle to return to the shock instead of 
being advected further downstream by the mean field \citep{Pelletier2009}. 
The latter condition can be conveniently written as
\begin{align}
\sigma \lesssim \sigma_{\rm F} \simeq \epsilon_B^2 (\lambda / d_{\rm i})^2 
(E/E_{0\rm i})^{-2},
\label{eq:fermi}
\end{align}
where $E_{0\rm i} = \gamma_0 m_{\rm i}c^2$.\footnote{The estimate \eqref{eq:fermi} is similar to the one
used in Sec.~\ref{sec:transition} to obtain $\sigma_{\rm L}$ for unmagnetized ions, 
but the question being asked is different. Here, we consider the near-downstream motion of the (nearly) 
isotropic suprathermal particle population, whereas Sec.~\ref{sec:transition} concerns the transport of the incoming background 
ions with typical energy $E\sim E_{\rm 0i}$ over the near precursor. We point out that 
the transport of the incoming background particles need not be, and generally is not, 
of the same nature as the transport of the suprathermal particle population (e.g., the background particles may 
be trapped in the Weibel filaments while the suprathermal population is not).}
Here, we point out two important aspects. First, through the dependence of $\epsilon_B$ 
and $\lambda$ on $Z_\pm$ (see Sec.~\ref{sec:theory}), the 
maximum external magnetization that allows for Fermi acceleration becomes a 
function of the pair-loading factor. And secondly, if the electrons (and positrons) do 
\emph{not} reach equipartition with the ions, as is generically the case
for $Z_\pm \gtrsim 1$, then the condition for electron Fermi cycles 
is different from the one for the ions. In particular, condition
\eqref{eq:fermi} becomes less restrictive for the relatively cooler electrons near the thermal peak
due to the inverse square dependence of $\sigma_{\rm F}$ on $E$, as long as this energy is high enough for 
a particle to remain untrapped (see discussion above).

Using \eqref{eq:eps_shock} and \eqref{eq:lambda_shock}, the condition \eqref{eq:fermi} can be expressed for
ions as
\begin{align}
\sigma \lesssim \sigma_{\rm Fi} 
\simeq \xi_{\rm b}^{3}\epsilon_{\rm e}^2(Z_\pm + 1)^{-4}\left(E/\overline E_{\rm i}\right)^{-2},
\label{eq:critical_fermi_ions}
\end{align}
where $\overline{E}_{\rm i}\sim E_{\rm 0i}$.
Taking typical values $\xi_{\rm b}\approx 0.1$ and $\epsilon_{\rm e}\approx 0.3$,
we find that nonthermal ions with energies in excess of the thermal 
component ($E\gtrsim \overline E_{\rm i}$) can be produced when
$\sigma\lesssim \sigma_{\rm Fi}  \sim 10^{-4}\times (Z_\pm + 1)^{-4}$. 
On the other hand, for electrons we obtain
\begin{align}
\sigma \lesssim \sigma_{\rm Fe} \simeq \xi_{\rm b}^{3}(Z_\pm + 1)^{-2}
\left(E/\overline E_{\rm e}\right)^{-2},
\label{eq:critical_fermi_el}
\end{align}
with $\overline E_{\rm e} \sim \epsilon_{\rm e}(Z_\pm + 1)^{-1} E_{0\rm i}$. 
Taking electrons with 
energy $E\sim \overline E_{\rm e}/\epsilon_{\rm e}$, we estimate
$\sigma_{\rm Fe} \sim 10^{-4}\times (Z_\pm +1)^{-2}$.

\begin{figure*}[htb!]
\centering
\includegraphics[width=\textwidth]{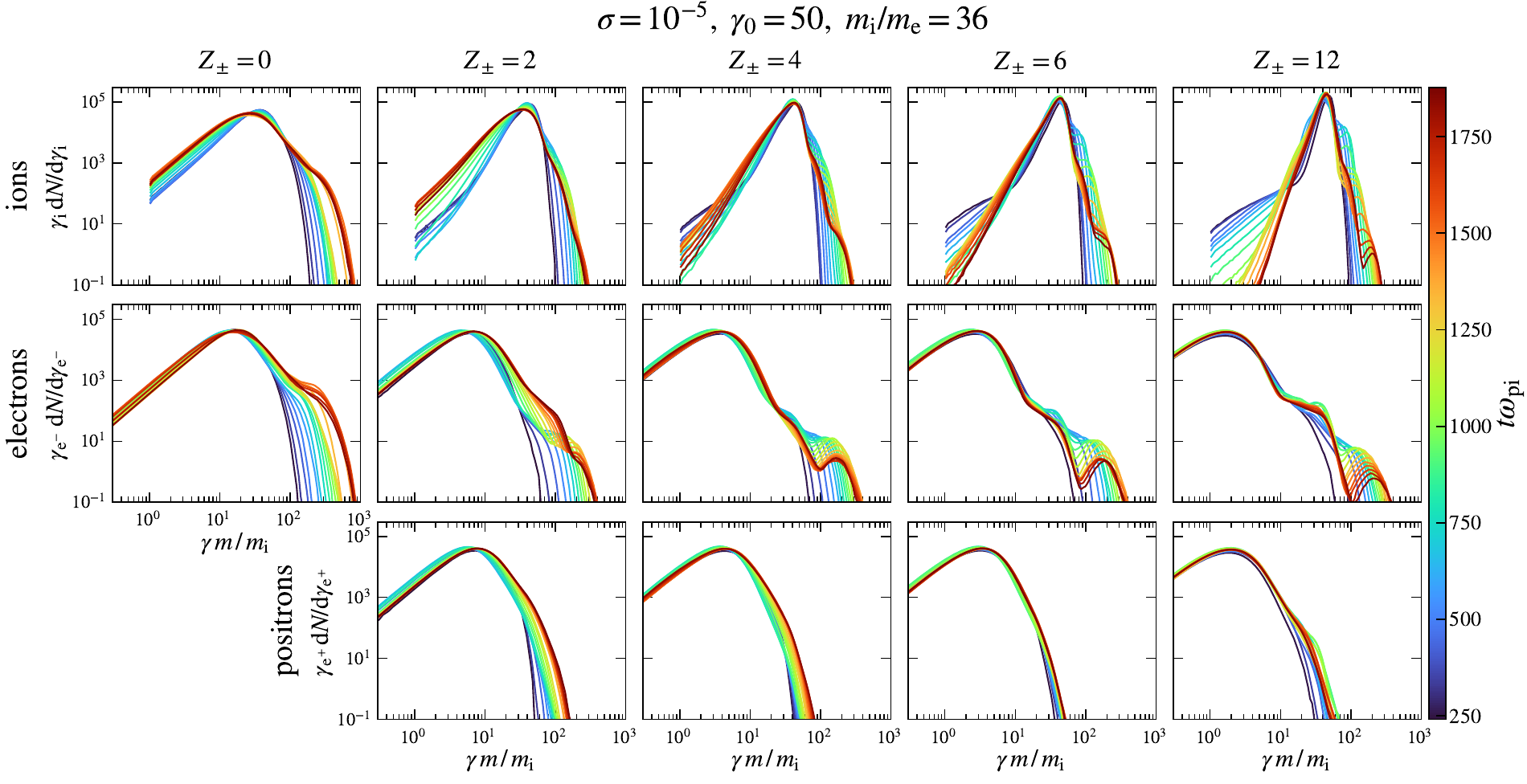}
\caption{\label{fig:spectra_sig1e-5}Energy spectrum evolution at fixed $\sigma = 10^{-5}$ and 
for increasing $Z_\pm$ (left to right) in a slice between $-150$ and $-100\,d_{\rm i}$ behind the ion shock.}
\end{figure*}

Based on the above, we envision a situation where nonthermal ion acceleration at the weakly magnetized shock is suppressed 
for $Z_\pm\gtrsim 1$, unless $\sigma$ is extremely low, such that $\sigma\lesssim \sigma_{\rm Fi}$. 
For electrons, limited acceleration remains possible as long as $\sigma\lesssim \sigma_{\rm Fe}$, even when the ions are thermal.
However, as soon as an electron is accelerated to energies of the order of $E\sim E_{0\rm i}$, 
the nature of its transport becomes similar to that of the thermal ions, 
implying that any nonthermal acceleration beyond $E\sim E_{0\rm i}$ is inhibited. Therefore, 
the electrons may form a nonthermal component even when the ions are essentially thermal, 
but the extent of the nonthermal tail will be in this case limited between 
$E\sim (Z_\pm + 1)^{-1}E_{0\rm i}$ and $E\sim E_{0\rm i}$. Finally, following the simplified physics picture discussed above, 
we expect electrons to be thermal when $\sigma\gtrsim \sigma_{\rm Fe}$ (although, see Fig.~\ref{fig:spectra_sig1e-5}).

We now compare the above estimates with results from PIC simulations. Fig.~\ref{fig:spectra_Z2} shows the 
evolution of the downstream particle energy spectrum at fixed $Z_\pm = 2$ and for different 
magnetizations $\sigma$.
The results shown represent some of our longest-duration simulations (in $\omega_{\rm pi}^{-1}$ units) 
and are as such best suited for probing the nature of particle acceleration. At the time when each 
$Z_\pm = 2$ simulation ends, we also show the spectra obtained for $Z_\pm = 0$ at the same time.
In line with the above discussion, we find that even a single electron-positron pair per ion is 
enough to suppress ion acceleration at magnetizations as low as $\sigma = 5\times 10^{-6}$. 
In contrast, 
the electron-ion shock produces nonthermal ions up to $\sigma = 3\times 10^{-5}$ \citep[see also][]{Sironi2013}.
These results are consistent with the estimate \eqref{eq:critical_fermi_ions}, which gives 
$\sigma_{\rm Fi}\approx 10^{-4},\, 10^{-6}$
for $Z_\pm = 0,\,2$, respectively. For electrons, condition \eqref{eq:critical_fermi_el} gives 
$\sigma_{\rm Fe}\approx 10^{-5}$ at $Z_\pm = 2$, and indeed we observe the development of a 
limited nonthermal electron tail in simulations 
with $\sigma = 10^{-5},\, 5\times 10^{-6}$ at $Z_\pm = 2$. For $\sigma = 3\times 10^{-5}$, the nonthermal
electron component (except for a minor kink in the spectrum) is only a transient, connected to the initial
reflection of particles from 
the wall on the left of the simulation domain (see also discussion of Fig.~\ref{fig:spectra_sig1e-5}).

\begin{figure}[htb!]
\centering
\includegraphics[width=\columnwidth]{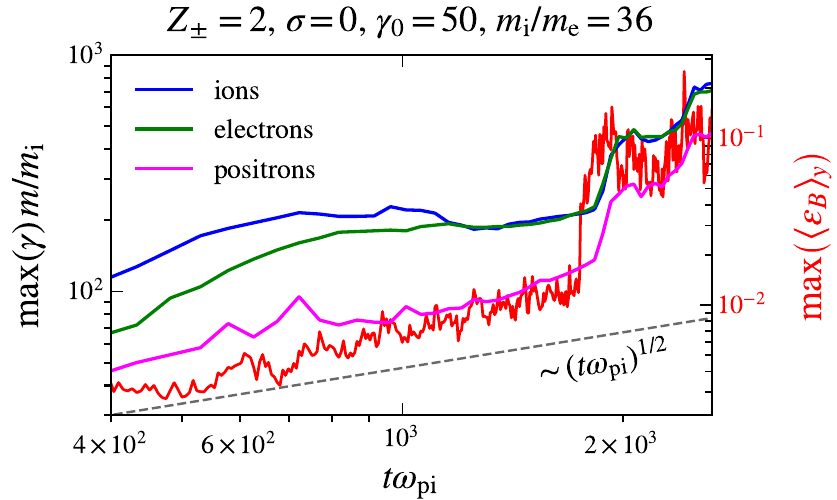}
\caption{\label{fig:max_ene} The maximum particle energy and magnetic energy fraction versus time 
in a $\sigma=0$ shock with moderate pair-loading factor. We determine $\max(\gamma)$ based on 
where $\gamma{\rm d}N/{\rm d}\gamma = 0.1$ in the selected units of Fig.~\ref{fig:spectra_Z2}. The maximum 
of $\langle\epsilon_B\rangle_y$ is 
measured at the shock in a $50\,d_{\rm i}$ wide slice. The scaling $\sim(t\omega_{\rm pi})^{1/2}$ 
is shown for reference only.}
\end{figure}

Unlike in the regimes with weak but finite $\sigma$, both ions and electrons form distinctly nonthermal distributions in the
limit of vanishing $\sigma$ (Fig.~\ref{fig:spectra_Z2}, rightmost panels). The acceleration is 
intermittent in time and correlated with the formation of
the magnetic cavities (see Sec.~\ref{sec:unmagnetized}). To demonstrate the point, we show in Fig.~\ref{fig:max_ene}
the evolution of the maximum particle energy (i.e., the spectral cutoff) behind the shock together with 
the maximum of the $y$-averaged magnetic energy fraction, measured in a slab around the shock.
Around the time $t\omega_{\rm pi}\approx$ 1700, when the first large-scale cavities appear in the near precursor, 
the magnetic energy fraction at the shock is amplified by an order of magnitude, followed by a rapid 
growth of the maximum particle energy. At select times, the maximum energy grows at a rate considerably faster
than $\max(\gamma)\sim (t\omega_{\rm pi})^{1/2}$, previously reported for 
electron-ion and pair plasma compositions \citep{Sironi2013,Plotnikov2018}. While the observed trend 
is intriguing, we note that longer duration simulations are required for a reliable extrapolation to astrophysically
relevant time scales.

\begin{figure}[htb!]
\centering
\includegraphics[width=\columnwidth]{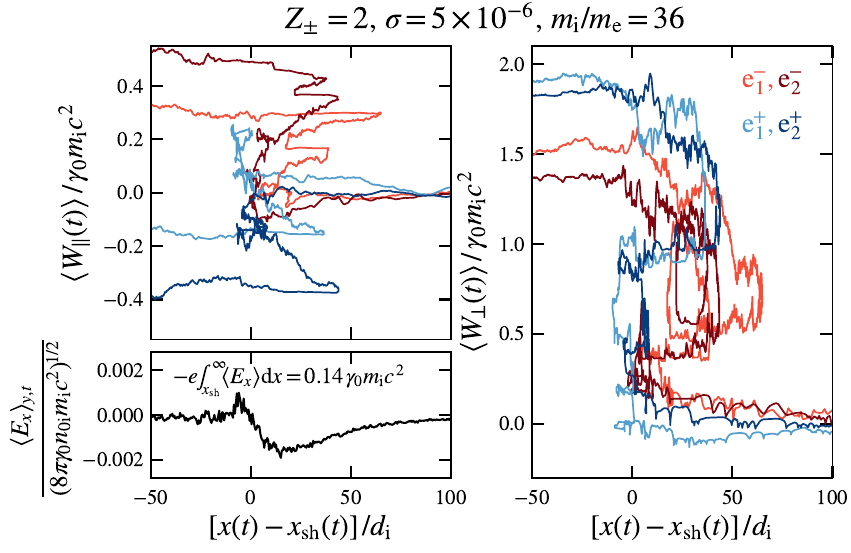}
\caption{\label{fig:acc_tracks} Work by 
the longitudinal ($E_x$) and transverse ($E_y$) 
electric field 
on a few high-energy positrons (blue lines) and electrons (red lines). Bottom left
panel shows the mean profile of $E_x(x- x_{\rm sh}(t),y,t)$, averaged over $y$ and $t$ (time interval matches the 
duration of particle tracking, $\Delta t\omega_{\rm pi}\approx 1000$).}
\end{figure}

The results of PIC simulations shown in Fig.~\ref{fig:spectra_Z2} reveal also 
a strong asymmetry between the electron and 
positron energy spectra. The highest energy electrons are accelerated to near 
equipartition with the highest energy ions, whereas the positrons are not. 
Moreover, the nonthermal component of the electron spectrum is much harder.
To explain the origin of the asymmetry, we consider in Fig.~\ref{fig:acc_tracks} the work by 
the $E_x$ and $E_y$ electric fields on a few representative high-energy electrons 
and positrons, extracted from the simulation with $\sigma=5\times 10^{-6}$ and $Z_\pm = 2$.
The work by the shock-perpendicular $E_y$ field is qualitatively similar for the two species
and exhibits random kicks in the particle energy that are characteristic of diffusive 
shock acceleration. On the other hand, the work by $E_x$ is largely 
mediated by a coherent field component (Fig.~\ref{fig:acc_tracks}, bottom left panel) that points toward 
the shock in the near upstream, within a distance of about 
$\sim 50\, d_{\rm i}$ ahead of the shock. 

On each cycle between the upstream and downstream, 
the returning electrons interacting with $E_x$ increase their energy by roughly 
$\Delta E / E_{0\rm i} \sim 0.1$, while the positrons lose an 
equivalent amount. This value is consistent with the estimated drop of the
electric potential from the shock toward the far upstream, 
which gives $e\int_{x_{\rm sh}}^{\infty}\langle E_x\rangle {\rm d}x \approx - 0.14\,E_{0\rm i}$.
The return of the particle from the upstream back to the shock plays little role in this energy exchange,
because most of the upstream residence time is spent by the particle moving away from 
the shock; when the particle turns around it is caught up by the shock rapidly.
Thus, the energy difference in the work done by $E_x$ accumulates 
upon repeated cycles, thereby favoring electron over positron acceleration. 
In the presence of magnetic cavities, some of the incoming background 
electrons (but not positrons) are 
preaccelerated near the cavities (see the longitudinal electron phasespace in Fig.~\ref{fig:unmagnetized}), 
which promotes the asymmetry further.

It is worth commenting on how the coherent $E_x$ field that favors 
electron over positron acceleration is generated. This field can be attributed to the fact that 
the returning beam ions carry on average higher relativistic inertia than the pairs, and therefore they 
penetrate further into the upstream, leaving behind most of the electrons and positrons with 
an excess negative charge. 
The resulting electrostatic potential gives rise to a near-upstream electric field that 
points in the negative $x$ direction. It should be mentioned that this coherent field is much smaller than the 
fluctuating fields near the shock transition (see Fig.~\ref{fig:den_filaments}). However, 
because it systematically affects the 
electron and positron energy gain on each Fermi cycle it leads to an overall significant difference 
between the nonthermal spectra of electrons and positrons.

For reference, we show in Fig.~\ref{fig:spectra_sig1e-5} the downstream energy spectrum evolution 
at fixed $\sigma=10^{-5}$ and for various $Z_\pm$.
We caution the reader that
beyond $Z_\pm$ of a few finite mass ratio effects for our choice of $m_{\rm i} / m_{\rm e}=36$ are not 
to be ruled out (see Appendix~\ref{app:mass_ratio}). With this caveat in mind, we report the following.  
Except when $Z_\pm = 0$, the ions are essentially thermal, 
lacking any substantial nonthermal component. The relatively cooler electrons 
with mean energy $\overline E_{\rm e} = \epsilon_{\rm e}(Z_\pm + 1)^{-1}E_{0\rm i}$
develop a limited nonthermal tail with a cuttof energy 
$E\sim (Z_\pm + 1)^{-1}E_{0\rm i}$. 
The acceleration of positrons
is disfavored by the mean electric field in front of the shock (see above discussion), 
such that the positrons remain nearly thermal.
The electron spectrum features as well a 
high-energy spectral bump that gradually recedes with time. Particle tracking (not shown) relates 
the high-energy bump with a transient energization of electrons near the tip of 
the particle precursor. The feature is therefore 
a remnant of the initial reflection of plasma from the simulation wall, and does 
not persist in the long-time regime of particle acceleration.  To summarize, the generic property 
that emerges from our simulations
of weakly magnetized pair-loaded shocks is that the ions are essentially thermal, 
whereas the electrons form a nonthermal tail of limited extent.

\begin{figure}[htb!]
\centering
\includegraphics[width=0.85\columnwidth]{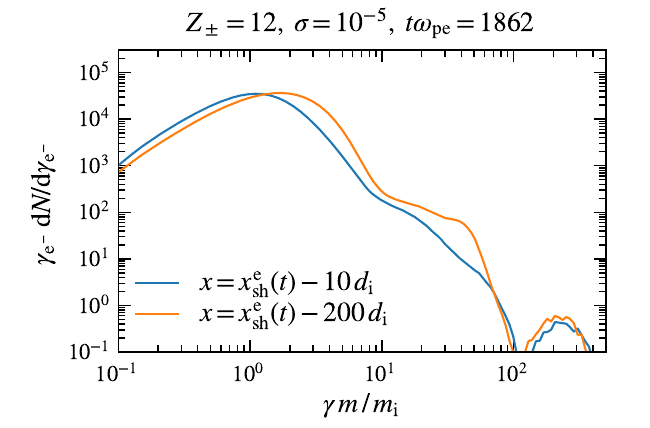}
\caption{\label{fig:spec_Z12_x} Electron spectrum in a 10 $d_{\rm i}$ wide slice at two different
$x$ locations behind the pair shock for $Z_\pm = 12$ and $\sigma = 10^{-5}$.}
\end{figure}

For $Z_\pm\geq 4$, the magnetization of $\sigma = 10^{-5}$ 
is above the estimated range of $\sigma$ (see Eq.~\eqref{eq:critical_fermi_el}) that allows for 
electron Fermi cycles. An important aspect to consider here is that for high $Z_\pm$ the electron shock lies 
ahead of the broader ion shock (Fig.~\ref{fig:den_prof}), and the space in between
is filled with a turbulent field sheared by a transverse $E_x \times B_z$ flow (see discussion of Fig.~\ref{fig:tracks}). 
This feature departs from the context in which the estimate \eqref{eq:critical_fermi_el} is made, 
where the particle 
scattering centers behind the shock are essentially at rest in the downstream frame. 
That significant electron energization indeed occurs behind their shock can be seen by 
inspecting the spectrum at different $x$ locations, as shown in Fig.~\ref{fig:spec_Z12_x}. 
The electron spectrum immediately behind the electron-positron 
shock exhibits a much softer nonthermal component
compared to the far downstream spectrum, behind the broader ion shock. Moreover, 
not only the high-energy component, but also the core of the particle distribution
is energized during the passage through the turbulent sheared layer in between the
two shocks.

\section{Astrophysical implications}
\label{sec:astro}

The results presented in this paper are relevant for the early phase of the GRB afterglow, when the
external shock propagates into a medium enriched with electron-positron pairs.
We provide direct estimates for the fraction of energy carried by the post-shock pairs and 
constrain the maximum external magnetization that allows for efficient particle acceleration.

GRB explosions may occur in the interstellar medium or inside the wind of a massive progenitor star, 
in particular of Wolf-Rayet type \citep{Crowther2007}. The magnetization of the 
interstellar medium is extremely low; it varies around $\sigma\sim 10^{-9}$. 
The magnetization of a Wolf-Rayet wind before the explosion is likely much higher than the 
magnetization of the interstellar medium, but its exact value is poorly known. 
An upper limit may be estimated using the wind kinetic energy per particle. 
This gives $\sigma\lesssim (w/c)^2 \sim 10^{-5}$ for typical wind velocities $w\sim 10^8\,{\rm cm/s}$.

The model of the early GRB afterglow developed by \citet{Beloborodov2014} 
shows good agreement with 
a set of GRB observations, 
assuming ambient densities typical of Wolf-Rayet type progenitors 
and emission from essentially thermal pairs behind the shock, carrying an energy fraction
$\epsilon_{\rm e}\approx 0.3$ when $1\lesssim Z_\pm \ll m_{\rm i}/m_{\rm e}$ \citep[see also][]{Hascoet2015}. 
Our first principles kinetic simulations support these assumptions. 
We find $0.2\lesssim \epsilon_{\rm e}\lesssim 0.5$ and rather limited nonthermal 
electron acceleration for magnetizations near the estimated upper limit of 
Wolf-Rayet stellar winds ($\sigma\sim 10^{-5}$). In this case, the maximum (downstream frame) 
energy of the nonthermal electrons 
is set by the shock Lorentz factor $\sim\gamma_0$ and the 
ion mass as $E\sim E_{0\rm i} = \gamma_0 m_{\rm i}c^2$. Efficient electron
acceleration beyond $E\sim E_{0\rm i}$ then requires either very small amounts of pair loading, 
expected at radii $R > R_\pm \sim 10^{17}$\,cm \citep{Beloborodov2002}, or extremely 
low magnetizations, such as those expected for the interstellar medium ($\sigma\sim 10^{-9}$).

\section{Summary and conclusions}
\label{sec:conclusion}

In this work, we study the microphysics of pair-loaded, weakly magnetized relativistic shocks using 2D kinetic PIC simulations. 
Our simulations focus on the regime of 
moderate pair-loading factors $Z_\pm\lesssim 10$, where the far upstream energy is dominated by ions.
We find the following:
\begin{enumerate}
    \item Pair loading decreases the strength and scale of the 
    self-generated turbulence over the weakly magnetized precursor, 
    leading to a reduced efficiency of particle scattering. We attribute 
    this effect to the screening of ion current filaments by the background pairs (Sec.~\ref{sec:theory}). 
    \item When the external magnetization exceeds a critical value $\sigma_{\rm L}$, 
    the shock becomes mediated by the gyration of ions in the background compressed 
    mean magnetic field (Secs.~\ref{sec:tracks} and \ref{sec:transition}). 
    This critical value decreases with $Z_\pm$, owing to
    the weakening of the self-generated turbulence, which mediates the shock for $\sigma\lesssim\sigma_{\rm L}$.
    \item The energy fraction $\epsilon_{\rm e}$, carried 
    by the post-shock pairs, is robustly in the range between 20\% and 50\% of the upstream ion energy (Sec.~\ref{sec:heating}). These values are 
    favored by models of the early GRB afterglow that account for the pair loading \citep[e.g.,][]{Beloborodov2014, Hascoet2015}. 
    The mean electron energy scales 
    as $\overline E_{\rm e}\simeq \epsilon_{\rm e}(Z_\pm + 1)^{-1}E_{0\rm i}$, where
    $0.2\lesssim \epsilon_{\rm e}\lesssim 0.5$ and $E_{0\rm i} = \gamma_0m_{\rm i}c^2$ is 
    the far upstream ion energy.
    \item Pair loading tends to inhibit nonthermal particle acceleration, most 
    notably for ions (Sec.~\ref{sec:acc}).
    We estimate that acceleration via the first-order Fermi process is possible only when 
    the external magnetization is below a critical, pair-loading-dependent value 
    $\sigma_{\rm F}\sim 10^{-4}\times(Z_\pm + 1)^{-4}(E/E_{0\rm i})^{-2}$, where $E$ is the energy of the injected particle. 
    Simulations indeed show that the ions are essentially thermal at magnetizations as low 
    as $\sigma\approx 5\times10^{-6}$, even when the plasma is loaded 
    with only single electron-positron pair per ion. 
    The electrons, 
    on the other hand, form a nonthermal component of limited extent in 
    the range between $E\sim(Z_\pm + 1)^{-1}E_{0\rm i}$ and $E_{0\rm i}$.
    \item The limit of vanishing external magnetization is different from the regime with 
    weak but finite $\sigma$ (Sec.~\ref{sec:unmagnetized}). 
    When $\sigma=0$, the microturbulence shows no apparent signs of weakening with
    growing $Z_\pm$; at least not for the order-unity values of $Z_\pm$ considered in our simulations. The
    locally intense fields are supplied by magnetized plasma cavities, generated over the turbulent
    precursor. Then, particle acceleration of 
    both ions and electrons is sustained over the duration of the entire simulation. We estimate that, 
    under the most favorable conditions, the external magnetization should be no larger than $\sigma\sim 10^{-6}$ 
    for the pair-loaded shock to be an efficient accelerator (Sec.~\ref{sec:unmag_disc}).
\end{enumerate}

The subject offers a number of promising future directions. In our setup, the pair enrichment is characterized 
by a single parameter, the pair-loading factor $Z_\pm$, neglecting the fact that 
the pairs are injected with a finite momentum in the rest frame of the external medium. The 
available free energy of the drifting pairs is released through plasma streaming instabilities, which preamplify 
magnetic fields in the far upstream \citep{Ramirez2007, Garasev2016, Derishev2016,Peterson2021b}. 
If these fields manage to survive until they are caught up by the shock, the
scale and strength of the fluctuations at the shock and further downstream could be modified.
Pair enrichment also plays a significant role in relativistic radiation mediated shocks, 
although the physics in that case is somewhat different from the regime consider here, 
owing to direct momentum exchange between the radiation and the plasma \citep[e.g.,][]{Levinson2020,Vanthiegem2022}. 
The present work motivates as well further studies along the lines of kinetic plasma theory. 
This includes, for instance, the theory for the slowdown of the background electrons and ions over the 
precursor of a pair-loaded shock, and the exact physical details required for the 
persistent generation of the magnetic cavities in low-$\sigma$ pair-loaded relativistic shocks.

\begin{acknowledgments}
We thank I.~\mbox{Plotnikov}, A.~\mbox{Philippov}, J.~\mbox{N\" attil\" a}, 
L.~\mbox{Comisso}, J.R.~\mbox{Peterson}, and A.~\mbox{Spitkovsky}
for helpful discussions related to this work.
The authors would like to acknowledge the OSIRIS Consortium, consisting of
UCLA and IST (Lisbon, Portugal), for the use of \textsc{osiris}
and for providing access to the \textsc{osiris 4.0} framework.
D.G.~was supported by the U.S.~DOE Fusion 
Energy Sciences Postdoctoral Research Program administered by ORISE for the DOE. ORISE is managed by 
ORAU under DOE contract number DE-SC0014664. All opinions expressed in this paper are the authors' and do not necessarily 
reflect the policies and views of DOE, ORAU, or ORISE. L.S.~acknowledges
support by NSF Grant No.~AST-1716567 and NASA Grant No.~ATP 80NSSC20K0565.
The authors acknowledge the Gauss Centre for Supercomputing e.V.~for 
funding this project by providing computing time on SuperMUC-NG at the Leibniz 
Supercomputing Centre under project No.~pr74vi (Principal Investigator: J\" org B\" uchner). 
Additional computing resources were also provided on NERSC Cori.
\end{acknowledgments}

\software{\textsc{osiris} \citep{Fonseca2002,Fonseca2013}}

\appendix
\section{Dependence on the mass ratio}
\label{app:mass_ratio}

\begin{figure}[htb!]
\centering
\includegraphics[width=0.85\columnwidth]{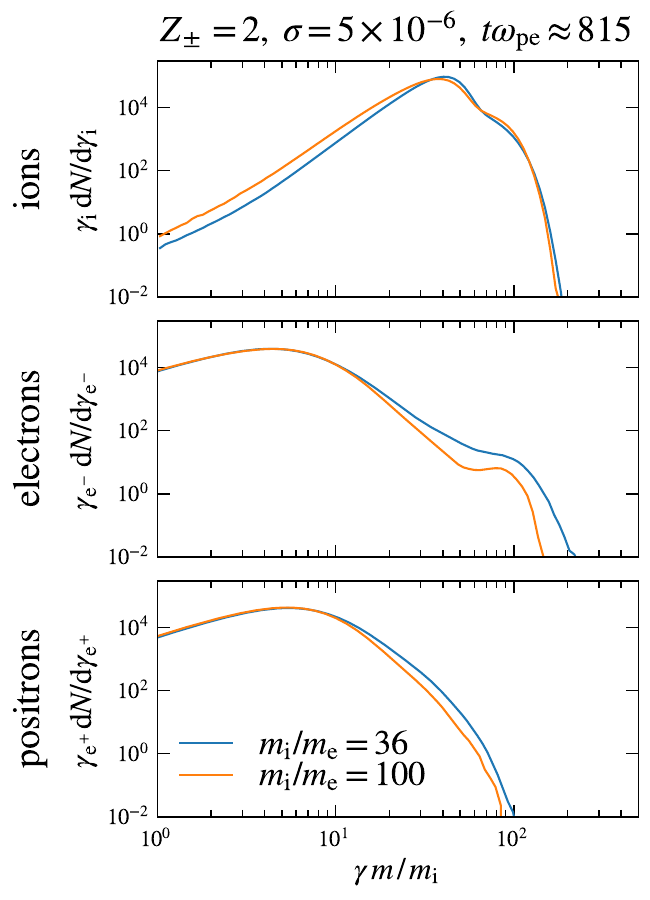}
\caption{\label{fig:spec_mi} Dependence of the post-shock particle energy spectrum 
on the ion-electron mass ratio. The spectra are measured in a slice between $-150$ and $-100\,d_{\rm i}$ behind the ion shock.}
\end{figure}
In Fig.~\ref{fig:spec_mi} we compare the downstream particle energy spectra around the time $t\omega_{\rm pi}\approx 815$ 
in a shock with $Z_\pm = 2$ and $\sigma = 5\times 10^{-6}$ for $m_{\rm i}/m_{\rm e}=36, 100$. To save resources, 
we perform the $m_{\rm i}/m_{\rm e}=100$ simulation using a $17\,d_{\rm i}$ wide box, with a resolution of 6.5 cells per $d_{\rm e}$ and 
four particles per cell per species. The $m_{\rm i}/m_{\rm e}=36$ simulation has a $31.4\,d_{\rm i}$ wide box,
with a resolution of eight cells per $d_{\rm e}$ and twelve particles per cell per species. 

As shown in Fig.~\ref{fig:spec_mi}, qualitatively and quantitatively similar results are
obtained at the increased value of the ion-electron mass ratio. We conclude that
the simulations from the main text using $m_{\rm i}/m_{\rm e}=36$  are reasonably 
converged in terms of the mass ratio 
for pair-loading factors up to a few.
This is consistent with \citet{Sironi2013}, 
who performed mass ratio scans (up to $m_{\rm i}/m_{\rm e} = 1600$) in simulations
of relativistic electron-ion shocks at $\sigma=10^{-5}$, and concluded that mass ratios 
as low as $m_{\rm i}/m_{\rm e} = 25$ are sufficient for reasonably converged results. It is also worth
highlighting the excellent agreement in the thermal parts of the electron and positron
spectra in Fig.~\ref{fig:spec_mi}, even though the far upstream pair energy fraction, 
$\epsilon_{\rm e0} = (Z_\pm + 1)m_{\rm e}/m_{\rm i}$, differs significantly between the two runs.

Computational limitations currently prevent long-duration shock simulations at mass ratios
much higher than 36 for $Z_\pm$ beyond a few. However, the following can be noted.
In Sec.~\ref{sec:heating} we demonstrate that the mean post-shock electron energy per particle
drops as $\overline{E}_{\rm e}\propto(Z_\pm + 1)^{-1}$ with the pair-loading factor.
The trend ceases when $\overline{E}_{\rm e}\approx \epsilon_{\rm e}(Z_\pm + 1)^{-1}\gamma_0 m_{\rm i}c^2 \approx \gamma_0 m_{\rm e}c^2$,
as the electrons get to keep their initial far upstream energy of
$\gamma_0 m_{\rm e}c^2$. Therefore, mass ratio effects should become significant
whenever $Z_\pm + 1\sim \epsilon_{\rm e} m_{\rm i}/m_{\rm e}$. This suggests that finite mass ratio 
effects could play a role in our simulations with $Z_\pm = 6, 12$ and $m_{\rm i}/m_{\rm e}=36$. 
On the other hand, it is worth noting that our results are supported by analytical estimates, 
derived under the general assumption $Z_\pm \ll m_{\rm i}/m_{\rm e}$ and without specifying any 
particular value for $m_{\rm i}/m_{\rm e}$.

\bibliography{refs}
\bibliographystyle{aasjournal}

\end{document}